\documentclass[a4paper,fleqn,usenatbib]{mnras}

\usepackage{newtxtext,newtxmath}
\usepackage[T1]{fontenc}
\usepackage{aecompl}
\usepackage{graphicx}	% Including figure files
\usepackage{amsmath}	% Advanced maths commands
\usepackage{amssymb}	% Extra maths symbols

\usepackage[table,xcdraw]{xcolor}
\usepackage{multirow}
\usepackage{comment} 
\usepackage{booktabs,caption,fixltx2e}
\usepackage[flushleft]{threeparttable}
\usepackage{rotating}
\usepackage{adjustbox}
\def\hii{H{\small \ II}\ }
\def\hii{\textsc{Hii} }

\newcommand{\as}{$^{\prime\prime}~$}
\newcommand{\am}{$^{\prime}~$}

\title[Galactic synchrotron emissivity distribution]
{Galactic synchrotron emissivity measurements between 250$^\circ$ $< l <$ 355$^\circ$ from the GLEAM survey with the MWA}

\author[H. Su et al.]{
H.~Su$^{1,2,3}$\thanks{E-mail: hongquan.su@icrar.org},
N.~Hurley-Walker$^{3}$,
C.~A.~Jackson$^{3}$,
N.~M.~McClure-Griffiths$^{4}$,
\newauthor
S.~J.~Tingay$^{3,5}$,
L.~Hindson$^{6,7}$,
P.~Hancock$^{3}$,
R.~B.~Wayth$^{3}$,
B.~M.~Gaensler$^{8,9,10}$,
\newauthor
L.~Staveley-Smith$^{11,9}$,
J.~Morgan$^{3}$,
M.~Johnston-Hollitt$^{7}$,
E.~Lenc$^{8,9}$,
M.~E.~Bell$^{12,9}$,
\newauthor
J.~R.~Callingham$^{8,9,12}$,
K.~S.~Dwarkanath$^{13}$,
B.-Q.~For$^{11}$,
A.~D.~Kapi\'{n}ska$^{11,9}$,
\newauthor
B.~McKinley$^{14}$,
A.~R.~Offringa$^{15}$,
P.~Procopio$^{14}$,
C.~Wu$^{11}$
and
Q.~Zheng$^{7}$
\\
$^{1}$~Key Laboratory of Optical Astronomy, National Astronomical Observatories, Chinese Academy of Sciences, Beijing 100012, China\\
$^{2}$~University of Chinese Academy of Science, 19A Yuquan Road, Beijing 100049, China\\
$^{3}$~International Centre for Radio Astronomy Research, Curtin University, Bentley, WA 6102, Australia\\
$^{4}$~Research School of Astronomy and Astrophysics, Australian National University, Canberra, ACT 2611, Australia\\
$^{5}$~Istituto di Radio Astronomia, Istituto Nazionale di Astrofisica, Bologna, Italy, 40123\\
$^{6}$~Centre for Astrophysics Research, School of Physics, Astronomy and Mathematics, University of Hertfordshire, \\ \ \ College Lane, Hatfield AL10 9AB, UK\\
$^{7}$~School of Chemical and Physical Sciences, Victoria University of Wellington, PO Box 600, Wellington 6140, New Zealand\\
$^{8}$~Sydney Institute for Astronomy, School of Physics, The University of Sydney, NSW 2006, Australia\\
$^{9}$~ARC Centre of Excellence for All-sky  Astrophysics (CAASTRO)\\
$^{10}$~Dunlap Institute for Astronomy and Astrophysics, University of Toronto, 50 St.\ George Street, Toronto, ON M5S 3H4, Canada\\
$^{11}$~ICRAR University of Western Australia, Crawley, WA 6009, Australia\\
$^{12}$~CSIRO Astronomy and Space Science (CASS), PO Box  76, Epping, NSW 1710, Australia\\
$^{13}$~Raman Research Institute, Bangalore 560080, India\\
$^{14}$~School of Physics, The University of Melbourne, Parkville, VIC 3010, Australia\\
$^{15}$~ASTRON, The Netherlands Institute for Radio Astronomy, Postbus 2, 7990 AA, Dwingeloo, The Netherlands
}

\date{Accepted XXX. Received YYY; in original form ZZZ}

\pubyear{2017}

\begin{document}
\label{firstpage}
\pagerange{\pageref{firstpage}--\pageref{lastpage}}
\maketitle

\begin{abstract}
Synchrotron emission pervades the Galactic plane at low radio frequencies, originating from cosmic ray electrons interacting with the Galactic magnetic field. Using a low-frequency radio telescope, the Murchison Widefield Array (MWA), we measure the free-free absorption of this Galactic synchrotron emission by intervening \hii regions along the line of sight. These absorption measurements allow us to calculate the Galactic cosmic-ray electron emissivity behind and in front of 47~detected \hii regions in the region $250^\circ < l < 355^\circ$, $|b| < 2^\circ$. We find that all average emissivities between the \hii regions and the Galactic edge along the line of sight ($\epsilon_b$) are in the range of 0.24$\,\,\sim\,\,$0.70$\,\,$K$\,\,$pc$^{-1}$ with a mean of 0.40$\,\,$K$\,\,$pc$^{-1}$ and a variance of 0.10$\,\,$K$\,\,$pc$^{-1}$ at 88$\,\,$MHz. Our best model, the Two-circle model, divides the Galactic disk into three regions using two circles centring on the Galactic centre. It shows a high emissivity region near the Galactic centre, a low emissivity region near the Galactic edge, and a medium emissivity region between these two regions, contrary to the trend found by previous studies.
%Our modelling shows high emissivity (8.2$^{+5.3}_{-3.5}\,\,$K$\,\,$pc$^{-1}$) near the Galactic centre within a Galactocentric radius of 1.7$^{+0.2}_{-0.3}\,\,$kpc, low emissivity (0.51$^{+0.19}_{-0.27}\,\,$K$\,\,$pc$^{-1}$) outside of a Galactocentric radius of 5.2$^{+4.7}_{-3.2}\,\,$kpc, and medium emissivity (0.84$^{+0.76}_{-0.55}\,\,$K$\,\,$pc$^{-1}$) between these two regions, contrary to the trend found by previous studies. 
\end{abstract}


%%%%%%%%%%%%%%%%%%%%%%%%%%%%%%%%%%%%%%%%%%%%%%%%%%

%%%%%%%%%%%%%%%%% BODY OF PAPER %%%%%%%%%%%%%%%%%%

%Figures are referred to as e.g. Fig.~\ref{fig:example_figure}, and tables as e.g. Table~\ref{tab:example_table}.

\section{Introduction}
The plane of the Milky Way is dominated by diffuse radio emission with a brightness temperature of thousands of Kelvin at low radio frequencies \citep{Zheng2016MNRAS.tmp.1510Z}. This emission originates from relativistic electrons interacting with the Galactic magnetic field. The total synchrotron power ($P$) radiated by one electron follows $P \propto E^2B^2$ where $E$ is the electron energy, and $B$ is the Galactic magnetic field strength \citep{Westfold1959ApJ...130..241W, Epstein1967ApJ...150L.109E}.
% $\sim$150$\,\,$MHz,

 Many previous low radio frequency observations have focused on two-dimensional (2-D) Galactic maps, i.e. tracing the distribution of brightness temperature with Galactic longitude and latitude. \citet{Oliveira-Costa2008MNRAS.388..247D} summarised 31 total power surveys in the frequency range $4\leq\nu\leq820$ MHz. All of these observations were performed between 1944 and 1982 with angular resolutions larger than several degrees. One widely-used all-sky map is that of \citet{Haslam1982A&AS...47....1H} at 408$\,\,$MHz with an angular resolution of 51$^{\prime}$. All of these 2-D maps show a bright Galactic plane with spatial emissivity\footnote{Emissivity is the brightness temperature per unit length with the unit of K~pc$^{-1}$ which is equivalent to 7.71$\,\,\times\,\,$10$^{-41}\,\,$W$\,\,$m$^{-3}\,\,$Hz$^{-1}\,\,$sr$^{-1}$ at 88$\,\,$MHz in our observations.} information integrated along the line of sight. Here we focus on the effect of the ionised hydrogen regions (\hii regions) on the synchrotron background that gives rise to absorption features against the bright diffuse synchrotron background. 
% There are also many surveys at other frequencies, but we focus on the low-frequency surveys related to the regions of ionised hydrogen (\hii region) absorption.
% At frequencies between 20 and 100$\,\,$GHz, the Planck and WMAP satellites observed the synchrotron emission in their polarised intensity maps \citep{Planck2015arXiv150606660P, Bennett2013ApJS..208...20B}. 

Absorption of synchrotron radiation by \hii regions provides an opportunity to study the emission distribution along the line of sight. This idea is not new, having been put forward in the 1950s \citep{Scheuer1953MNRAS.113....3S}. The diffuse synchrotron emission can be absorbed by the free electrons in \hii regions via free-free absorption. The optical depth of an \hii region is proportional to the inverse square of the frequency \citep{Condon1992ARA&A..30..575C}. At frequencies below about 200$\,\,$MHz, \hii regions become nearly opaque and show absorption features \citep{Mezger1967ApJ...147..471M, Kurtz2005IAUS..227..111K}. Thus, an \hii region separates the emission behind it and in front of it along the line of sight, when completed with information about the \hii region distance, enabling the measurement of the radial emissivity distribution with respect to the Sun.

To date, a modest number of \hii region absorption measurements have been made, limited by the poor angular resolution of telescopes observing at low radio frequencies. Most synchrotron emissivity calculations were made from samples each with less than 18 objects of sufficient angular resolution \citep{Jones1974AuJPh..27..687J, Caswell1976MNRAS.177..601C, Krymkin1978Ap&SS..58..347K, Abramenkov1990IAUS..140...49A, Fleishman1995A&A...293..565F, Roger1999A&AS..137....7R}. Prior to Nord et al. (2006, hereafter N06) only 46 emissivity measures had been collected in the literature. In the largest study to date, N06 observed the Galactic centre in the range of $-15^{\circ} < l < 26^{\circ}$, $|b| < 5^{\circ}$ at 74$\,\,$MHz using the Very Large Array (VLA). They detected 42~absorbing \hii regions (mainly in the range of $6^{\circ}\,\,<\,\,l\,\,<\,\,26^{\circ}$) with known distances from which they modelled the emissivity distribution and found the data to be consistent with a constant emissivity of 0.36$\,\,\pm\,\,$0.17$\,\,$K$\,\,$pc$^{-1}$ at 74$\,\,$MHz beyond 3$\,\,$kpc and zero emissivity within 3$\,\,$kpc of the Galactic centre.

% 3d models
The above observations require three-dimensional (3-D) modelling to reveal the distribution of the Galactic magnetic field and cosmic-ray electrons. There have been an ever improving set of models for one or both \citep{Beuermann1985A&A...153...17B, Han2006ApJ...642..868H, Sun2008A&A...477..573S, Sun2010RAA....10.1287S, Sun2012A&A...543A.127S, Orlando2013MNRAS.436.2127O}. The accuracy of any emissivity calculation is fundamentally tied to the quality of the underlying models. 

%In summary, the 2-D synchrotron emissions at different frequencies have been well observed in previous surveys. However, 
In total, emissivities have only been determined for $\sim$90 \hii regions across the entire Galactic plane and all of these values are critical to the accuracy of models for the distributions of the magnetic field and cosmic-ray electrons. In this work, which is the pilot for a larger study, we use the superb capabilities of the MWA to determine the emissivities towards 47 \hii regions, making this the largest single study of this kind to date. In Section~\ref{sec:data}, we introduce our observations and methods of emissivity measurements. In Section~\ref{sec:emi_measure}, we show our measured emissivities. In Section~\ref{sec:modelling}, we model the emissivity distribution. In Section~\ref{sec:sum}, we summarise our results and discuss future work.

\section{MWA 88$\,\,$MHz data and analysis}
\label{sec:data}

\subsection{Data}
 We use data from the GLEAM survey: the GaLactic and Extragalactic All-sky MWA survey \citep{Wayth2015PASA...32...25W}. The observations used in this study were performed for 10 hours on 17 March, 2014. The data reduction follows the method described in the MWA commissioning survey \citep{Hurley-Walker2014PASA...31...45H} with more specific details discussed in \citet{Hindson2016PASA...33...20H}. The main steps of data reduction include flagging, calibrating, and imaging individual snapshots. The calibrated Stokes XX and YY snapshots were imaged with a robust weighting of 0.0. To correct the astrometric changes in source position generated by the ionosphere \citep{Loi2015MNRAS.453.2731L}, we cross-matched the position of compact sources with MRC sources \citep{Large1981MNRAS.194..693L, Large1991Obs...111...72L} to determine an average astrometric correction and shifted each snapshot accordingly.	

\begin{figure*}
  \begin{adjustbox}{addcode={\begin{minipage}{\width}}{\caption{%
      A portion of the Galactic plane at 88$\,\,$MHz with a bandwidth of 30.72$\,\,$MHz observed by the MWA \citep{Hindson2016PASA...33...20H}. The angular resolution is 5\farcm6$\,\,\times\,\,$5\farcm6. The white polygons and yellow circles show the \hii regions with detected and undetected absorption features, respectively. Several absorption regions near the top-left corner in the top panel are detected but not measured due to the overlapped \hii regions with different distances. The striations in the middle of the second panel is caused by the bright Centaurus A. The black boxed region is enlarged in Fig.~\ref{fig:abs} to show an example absorption feature. Throughout we use the ``cubehelix" colour map \citep{Green2011BASI...39..289G}. The colour scale is adapted to ensure the Galactic features are highlighted.
   \label{fig:GP_img}
}\end{minipage}},rotate=90,center}
      \includegraphics[scale=0.7]{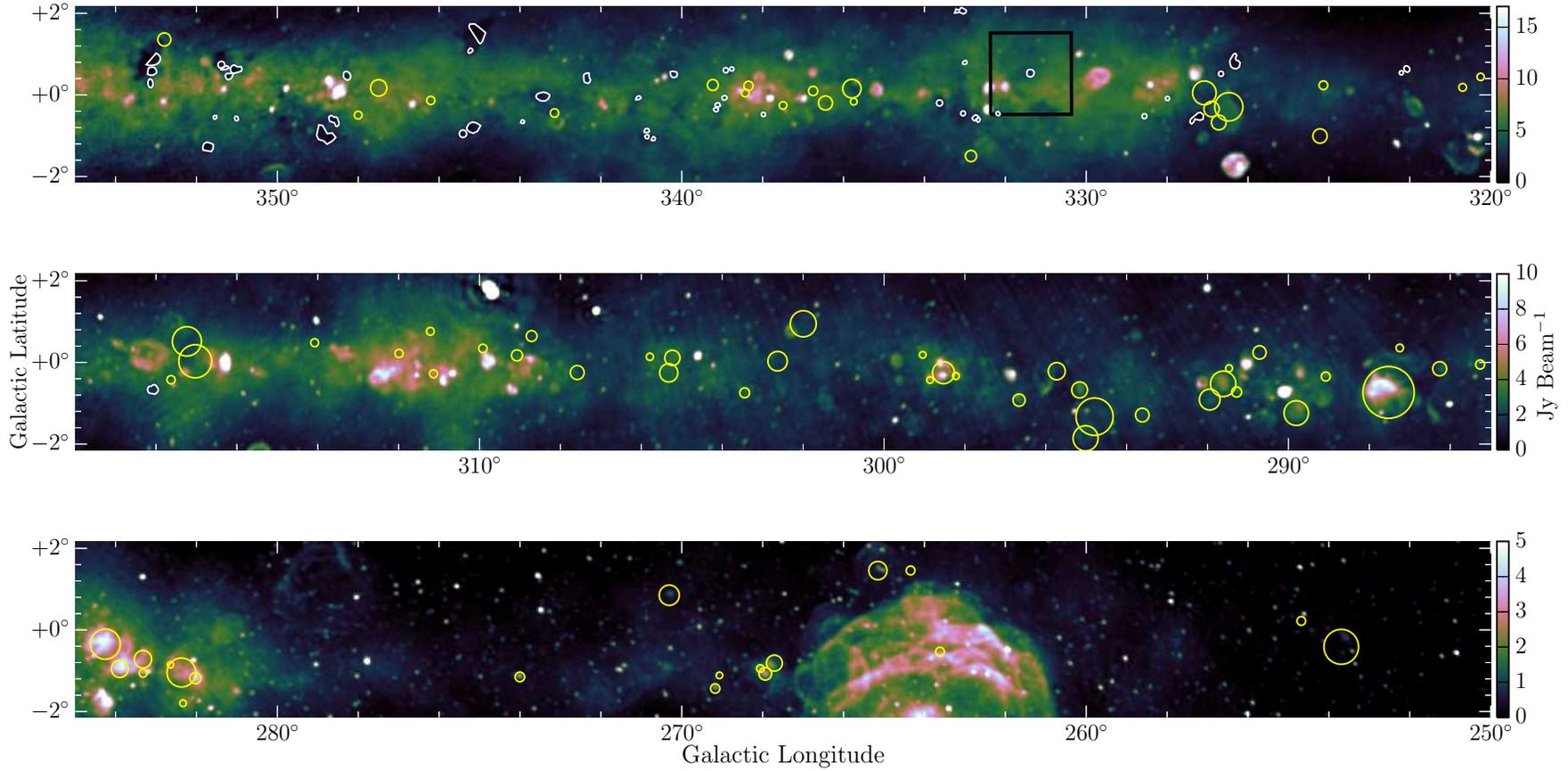}%
  \end{adjustbox}
\end{figure*}

The GLEAM data span 72 to 231$\,\,$MHz divided into five bands. The shortest baseline ($D_{\mathrm{min}}$) is 7.7$\,\,$metres corresponding to an angular scale of $25^{\circ}$ at 88$\,\,$MHz \citep{Tingay2013PASA...30....7T}. The surface brightness sensitivity ranges from 50 to 100$\,\,$mJy$\,\,$beam$^{-1}$ for the range of angular sizes from 5$^\circ$ to 15$^\circ$ \citep{Hindson2016PASA...33...20H}. In this paper, we use the data at the lowest frequency of 88$\,\,$MHz with a bandwidth of 30.72 MHz to maximize \hii region absorption detections. These data have a 1$\sigma$ sensitivity of 99$\,\,$mJy$\,\,$beam$^{-1}$ with an angular resolution of 5\farcm6$\,\,\times\,\,$5\farcm6 (1$\,\,$Jy$\,\,$beam$^{-1}\,\,$=$\,\,$1413.8$\,\,$K). The noise level is the lowest at Galactic longitude $\sim$300$^\circ$ and increases towards lower and higher Galactic longitudes shown in Fig.~\ref{fig:GP_img}.

Previous single dish surveys can recover the total power of the sky but their angular resolutions are in the order of degrees, which is not sufficient to resolve distant or small Galactic \hii regions. By contrast, the excellent angular resolution of the GLEAM survey enables the detection of \hii regions with minimum sizes from 2 to 33$\,\,$pc for distances from 1 to 20 kpc \citep{Hindson2016PASA...33...20H}. However, as an interferometer, the MWA is insensitive to power from large scale structures. The MWA consists of 128 aperture arrays distributed in a dense core $<$1.5$\,\,$km in diameter with a maximum baseline of about 2.5$\,\,$km. Thus, MWA has excellent $(u,v)$ coverage and surface brightness sensitivity to structures on angular scale from 5\farcm6 to 31$^\circ$ at 88$\,\,$MHz, which is sufficient for our emissivity measurements with \hii regions of typical angular scale sizes of several arcminutes. The observations of N06 are only sensitive to structures with angular sizes from $\sim$10\am to 40$^{\prime}$, resulting in negative temperature measurements for their observed absorption regions. Our observations also differ from N06 in spatial coverage. Our current measurements cover the southern sky in the range 250$^\circ <\,\,l\,\,<\,\,$355$^\circ$, $|b| < 2^\circ$, whereas N06 mainly covers the northern sky in the range $-$15$^\circ\,\,<\,\,l\,\,<\,\,26^\circ$, $|b| < 5^\circ$, furthermore the work of N06 was conducted at 74$\,\,$MHz, whereas the work in this paper is based on 88$\,\,$MHz. 

% \DIFaddbegin \DIFadd{The striations in the middle of the second panel is caused by the bright Centaurus A.} \DIFaddend

\subsection{Emissivity calculations}
\label{sec:formula}
\begin{figure}
\includegraphics[width=0.48\textwidth]{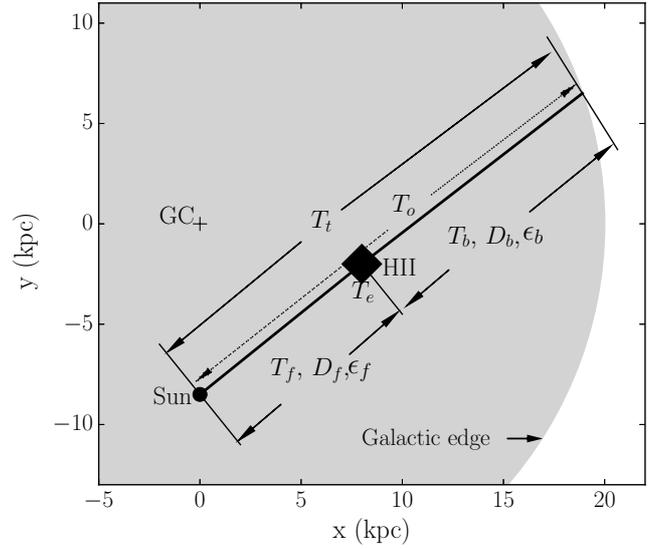}
\caption{A schematic diagram showing a portion of the Galactic disk (grey), the Galactic centre (GC), an \hii region, the Sun, the temperatures ($T_t,\,T_o,\,T_f,\,T_b,\,T_e$), the path lengths ($D_f,\,D_b$), and the emissivities ($\epsilon_f,\,\epsilon_b$).}
\label{fig:abs_cartoon} 
\end{figure}

 As discussed above, \hii regions become optically thick at low frequencies, absorbing the diffuse synchrotron radiation emitted behind them. An \hii region whose distance is known is a valuable probe of the synchrotron emission behind and in front of it. Here we introduce the formalism for using \hii regions to determine the average emissivity along the line of sight. The quantities of foreground and background emissivity are shown in Fig.~\ref{fig:abs_cartoon}.
 
Similar to a single dish telescope \citep{Kassim1987PhDT........10K}, we observe the \hii region with a brightness temperature of 
\begin{equation}
T_o = T_e(1-e^{-\tau}) + T_{b}e^{-\tau} + T_{f},
\label{eq:T_o}
\end{equation}
where $T_e$ is the electron temperature of the \hii region, $T_{b}$ is the brightness temperature of the synchrotron emission from the \hii region to the Galactic edge along the line of sight, $T_{f}$ is the brightness temperature of the synchrotron emission from the \hii region to the Sun, and $\tau$ is the optical depth of the \hii region. 

Usually, \hii regions become optically thick ($\tau \gg 1$) at frequencies below approximately 200$\,\,$MHz \citep{Mezger1967ApJ...147..471M}. So the observed brightness temperature of the \hii region becomes
\begin{equation}
T_o = T_e + T_{f}.
\label{eq:To}
\end{equation}
The brightness temperature of the sky neighbouring the \hii region $T_{t}$ is 
\begin{equation}
T_{t} = T_{f} + T_{b}.
\label{eq:Tt}
\end{equation}
Defining the average emissivity as the brightness temperature per unit length, the average emissivities behind ($\epsilon_{b}$) and in front ($\epsilon_{f}$) of \hii regions are
\begin{equation}
\begin{aligned}
& \epsilon_{f} = T_{f}/D_{f} = (T_o - T_e)/D_{f}, \\
& \epsilon_{b} = T_{b}/D_{b} = (T_t - T_o + T_e)/D_{b}
\end{aligned} 
\label{eq:emi}
\end{equation}
where $D_{f}$ is the distance between the \hii region and the Sun and $D_{b}$ is the distance between the \hii region and the edge of the Galactic plane. $T_e$ and $D_{f}$ are from the literature (see the references in Table~\ref{tab:emi}). $T_e$ is mainly measured from radio recombination lines at about 9$\,\,$GHz (\citealt{Balser2015ApJ...806..199B} and reference therein). For \hii regions without measured $T_e$, we estimate from the relation between the electron temperature and the Galactocentric radius $T_e\,\,=\,\,$(4928$\,\,\pm\,\,$277)$\,\,+\,\,$(385$\,\,\pm\,\,$29)$\,\,R_{\textrm{gal}}$ from \citealt{Balser2015ApJ...806..199B} (a similar relation is derived in \citealt{Alves2012MNRAS.422.2429A}). When estimating the electron temperature using the relation in \citet{Balser2015ApJ...806..199B} and \citet{Alves2012MNRAS.422.2429A} respectively, the differences are 1-8\% on $\epsilon_b$, depending on the distance of \hii regions. The small differences do not change our modelling results. $D_{f}$ is mainly measured using kinematics and parallax methods, which are summarised in \citet{Anderson2014ApJS..212....1A}. We calculate $D_{b}$ assuming a Galactocentric radius of 20$\,\,$kpc (\citealt{Ferriere2001RvMP...73.1031F} and references therein). Note that our $\epsilon_b$ depends on this assumed Galactocentric radius. $T_o$ and $T_t$ are from our observations. 

Unlike a single-dish telescope, an interferometer is incapable of measuring the total power of the Galactic plane emission. The data misses an unknown background offset across a large sky area, resulting in $\sim$20\% error in the absolute level of the measurements. As a result our measurements of $\epsilon_{f}$ have an additive offset or a scaling error and are not absolutely correct (Equation~\ref{eq:emi}). However, they are relatively correct because the missing power on scales less than a few degrees is absent equally from $T_t$ and $T_o$, so the difference $T_t - T_o$ in $\epsilon_b$ (Equation~\ref{eq:emi}) is correct. However, because $\epsilon_f$ relies only on $T_o$, which is not absolutely calibrated, we will not use $\epsilon_f$ in our modelling. 

It should be noted that the above formulae differ from those used by N06 because the MWA recovers the large-scale background emission to a scale of $\sim \lambda / D_{\mathrm{min}} = 25^\circ$ surrounding the \hii regions, whereas N06's observations do not. 

\subsection{H{\sc{ii}} region selection criteria}
\label{sec:rules}
In this paper we work with a sample of \hii regions with known distances and obvious absorption features to measure the average emissivities along a path. Using the 12 and 22$\,\,\mu$m data with high angular resolutions of 6\farcs5 and 12\as respectively from the Wide-field Infrared Survey Explorer (WISE), \citet{Anderson2014ApJS..212....1A} constructed a catalogue of $\sim$8000 Galactic \hii regions. This WISE \hii region catalogue constrains each \hii region within a radius. To select our sample, we search for absorption regions within these radii. We omitted overlapping \hii regions with different distances in our sample due to the complexity in obtaining reliable average emissivities. Note that \citet{Hindson2016PASA...33...20H} have presented a catalogue of 306 \hii regions using the data with all the frequencies between 72 and 231$\,\,$MHz from the MWA GLEAM survey. We do not use this catalogue because most of their measurements do not show obvious absorption features, and they did not publish the part of the sky closer to the Galactic Centre.

 We define absorption regions such that they are each: 
\begin{enumerate}
	\item inside the radius defined in the WISE \hii region catalogue; 
	\item lower than the nearby $T_t$ by at least 3$\sigma$ in surface brightness; 
	\item larger than the beam size to avoid beam dilution; 
	\item coincident with $12\,\,\mu$m emission features in WISE observations \citep{Wright2010AJ....140.1868W} given the 12$\,\,\mu$m emission is mainly from polycyclic aromatic hydrocarbon (PAH) molecules, which traces ionization fronts. The 22$\,\,\mu$m emission traces small dust grains composing the inner core of an \hii region (see \citealt{Deharveng2010A&A...523A...6D, Anderson2014ApJS..212....1A} and reference therein). Thus, we expect the 12$\,\,\mu$m emission region to be coincident with the absorption feature caused by free electrons from ionisation, whereas the 22$\,\,\mu$m emission region is much smaller than the absorption feature. 
\end{enumerate}

 For each absorption feature, we choose a nearby, non-absorbed area to estimate $T_t$ such that the area:
\begin{enumerate}
    \item is located in the similar Galactic latitude as the absorption region given the brightness temperature decreases rapidly with latitude;
    \item does not overlap with any \hii region in the WISE \hii region catalogue;
    \item does not overlap with any supernova remnants listed by \citet{Green2015ApJ...810...25G} to avoid contamination from bright non-thermal emission in our observations;
    \item does not have an obvious coherent structure, in order to avoid the effects caused by unknown sources, such as the undetected \hii regions and supernova remnants.
\end{enumerate}

\begin{figure*}
\includegraphics[width=0.51\textwidth]{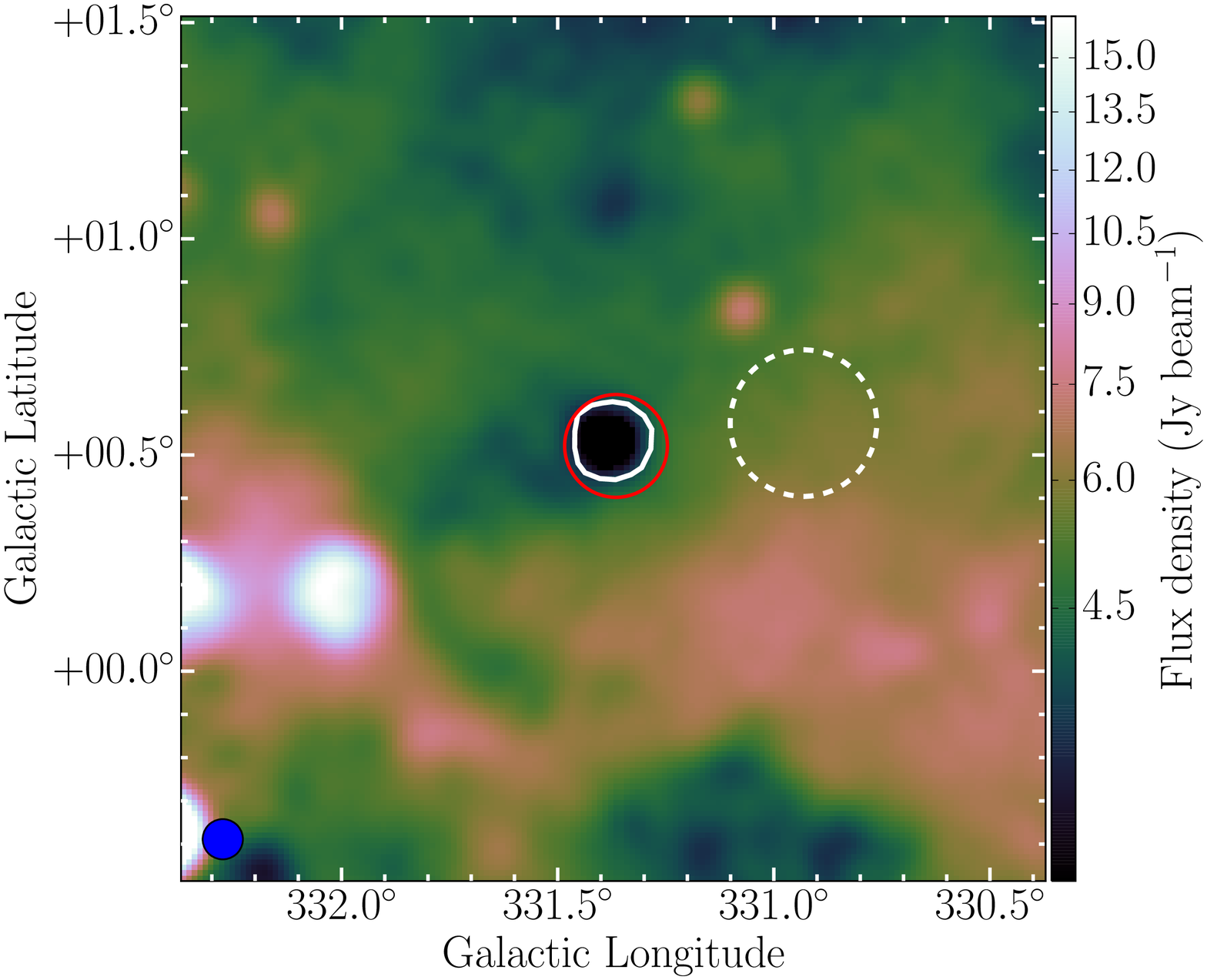}
\includegraphics[width=0.455\textwidth]{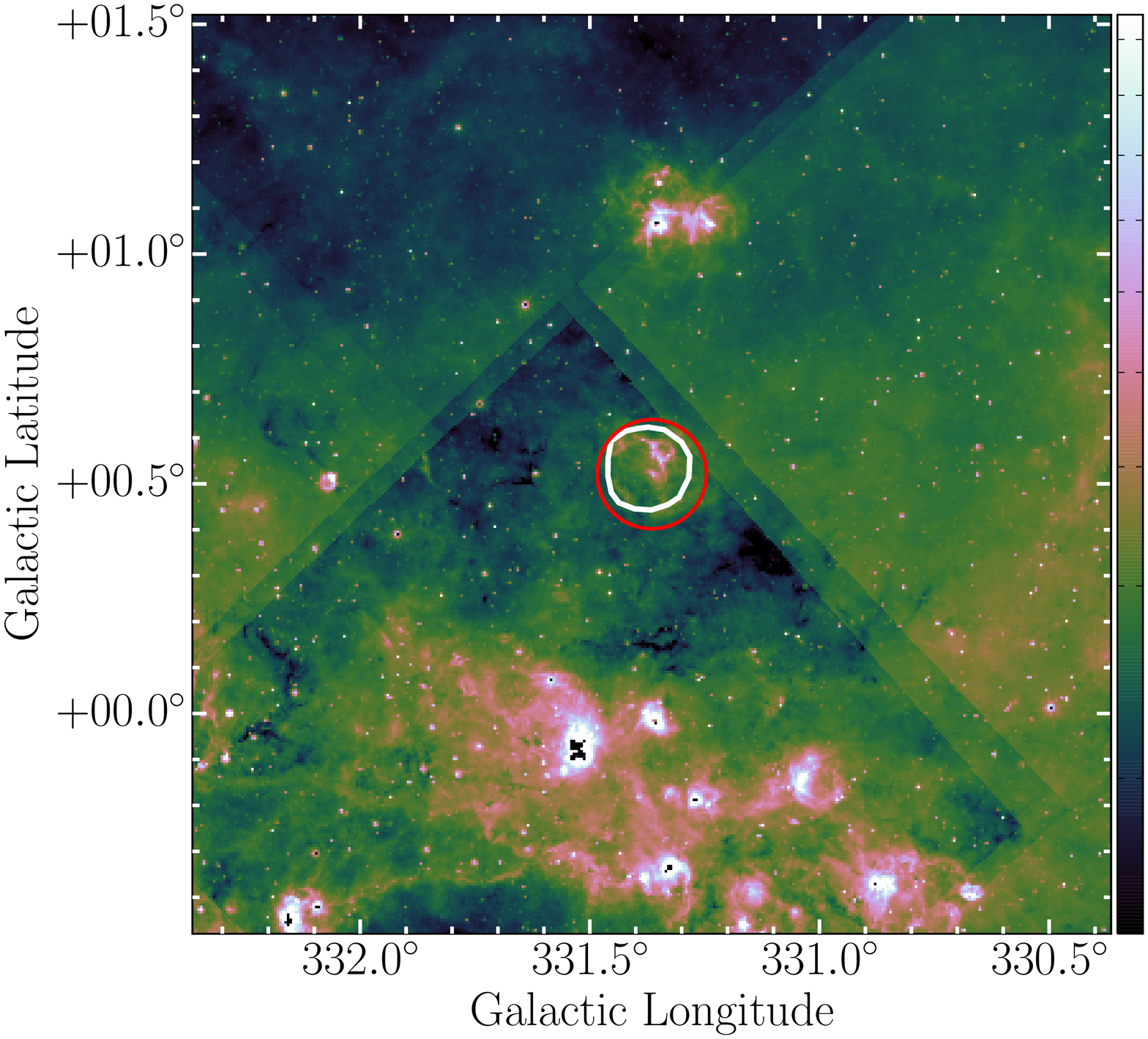}
\caption{An example of \hii region (G331.365+00.521) absorption feature at 88$\,\,$MHz from the MWA (left) and an emission feature at $12\,\,\mu$m from WISE (right). The white solid and dashed lines show our defined absorption and neighbouring regions respectively. See section~\ref{sec:rules} for the rules for selecting the absorbed and neighbouring regions. The red circle shows this \hii region defined in the WISE \hii region catalogue. The blue circle in the left corner shows the FWHM of the synthesised beam. The root-mean-square of this \hii region in this MWA image is about 0.26$\,\,$Jy$\,\,$beam$^{-1}$. This entire field is marked by a square in Fig.~\ref{fig:GP_img}.}
\label{fig:abs} 
\end{figure*}

We show an example of an absorption region and a neighbouring region in Fig.~\ref{fig:abs}. We confirm that the low surface brightness region in the centre of the MWA image (Fig.~\ref{fig:abs} left) is from \hii region absorption by comparing with the WISE image (Fig.~\ref{fig:abs} right). The WISE image is constructed by stacking 15 facets and then cropped to the same size as the MWA image. The line-like features are edges of image facets. These features do not have an effect on determining the location of the \hii region. 

\section{New emissivity measurements}
\label{sec:emi_measure}
\subsection{The emissivity distribution}
\label{sec:emi_distri}
 We use Equation~\ref{eq:emi} and the MWA measurements of $T_t$ and $T_o$ to estimate the average emissivities both behind and in front of 47~\hii regions (see Table~\ref{tab:emi}). We find that all average emissivities between the \hii regions and the Galactic edge along the line of sight ($\epsilon_b$) are in the range of 0.24$\,\,\sim\,\,$0.70$\,\,$K$\,\,$pc$^{-1}$ with an average of 0.40$\,\,$K$\,\,$pc$^{-1}$ and a variance of 0.10$\,\,$K$\,\,$pc$^{-1}$. The average emissivities between the \hii regions and the Sun ($\epsilon_f$) have a large range of -12.72$\,\,\sim\,\,$0.50$\,\,$K$\,\,$pc$^{-1}$ with a mean of -1.56$\,\,$K$\,\,$pc$^{-1}$ and a variance of 2.2$\,\,$K$\,\,$pc$^{-1}$, although these are subject to an uncertainty in total power affecting $\epsilon_f$. 
 
 We estimate the local root-mean-square (RMS) noise for the measured surface brightness ($I_o$ and $I_t$) using the standard deviation of pixel values in a representative neighbouring region. To incorporate the uncertainty in the distance measurement used in each \hii region we proceed as follows (i) where an error is explicitly stated in the literature this is used verbatim. Where no error has been quoted, then (ii) \hii regions with parallax distance measurements are assigned a 10\% error, (iii) \hii regions with a kinematic distance measurement are assigned a 50\% error and (iv) for all other cases, including \hii regions without a specified distance measurement method, we assign an error of 100\%. The errors of the electron temperature ($T_e$) of \hii regions are from the literature. Each above error is propagated to the average emissivities ($\epsilon_f$ and $\epsilon_b$). All the quoted errors are 1$\sigma$.  
 
The magnitude of any extragalactic synchrotron emission is negligible compared with Galactic synchrotron since its brightness temperature is in the range from 195 to 585$\,\,$K at 88$\,\,$MHz which is only $\sim\,\,$2\% of the Galactic component (\citealt{Guzman2011A&A...525A.138G} and references therein). Its effect ($\sim\,\,$0.03$\,\,$K$\,\,$pc$^{-1}$) on our measurements is smaller than our minimum error (0.05$\,\,$K$\,\,$pc$^{-1}$) and smaller than our average error (0.06$\,\,$K$\,\,$pc$^{-1}$).
 
 We give an indicative quality number to each measurement as a means of judging reliability (Table~\ref{tab:emi} Col.~10). The quality number is the number of matched conditions below, maximum value of 6 means all conditions met.
\begin{enumerate}
	\item The \hii region is isolated and does not overlap with other \hii regions.
	\item The \hii region belongs to a ``known" \hii region group, as defined in the WISE \hii region catalogue \citep{Anderson2014ApJS..212....1A}. ``Known" \hii regions are associated with radio recombination lines or H$\alpha$ emission; they are confirmed \hii regions, unlike ``group" and ``candidate" \hii regions.
	\item The area of the absorption region is at least two times larger than the beam size.
	\item The location of the absorption feature corresponds well with that of the WISE 12$\,\,\mu$m emission feature \citep{Wright2010AJ....140.1868W}. We use this criterion because the lowest temperature region in some absorption features has a positional or dimensional offset compared with the brightest emission feature in WISE. 
	\item The distance of the particular \hii region is measured rather than assuming it has the same distance as any other associated \hii regions.
	\item The electron temperature of the \hii region is measured rather than calculated from the statistical relation in \citet{Balser2015ApJ...806..199B}. 
\end{enumerate}  
 
\begin{figure*}
\centering
\includegraphics[width=\textwidth]{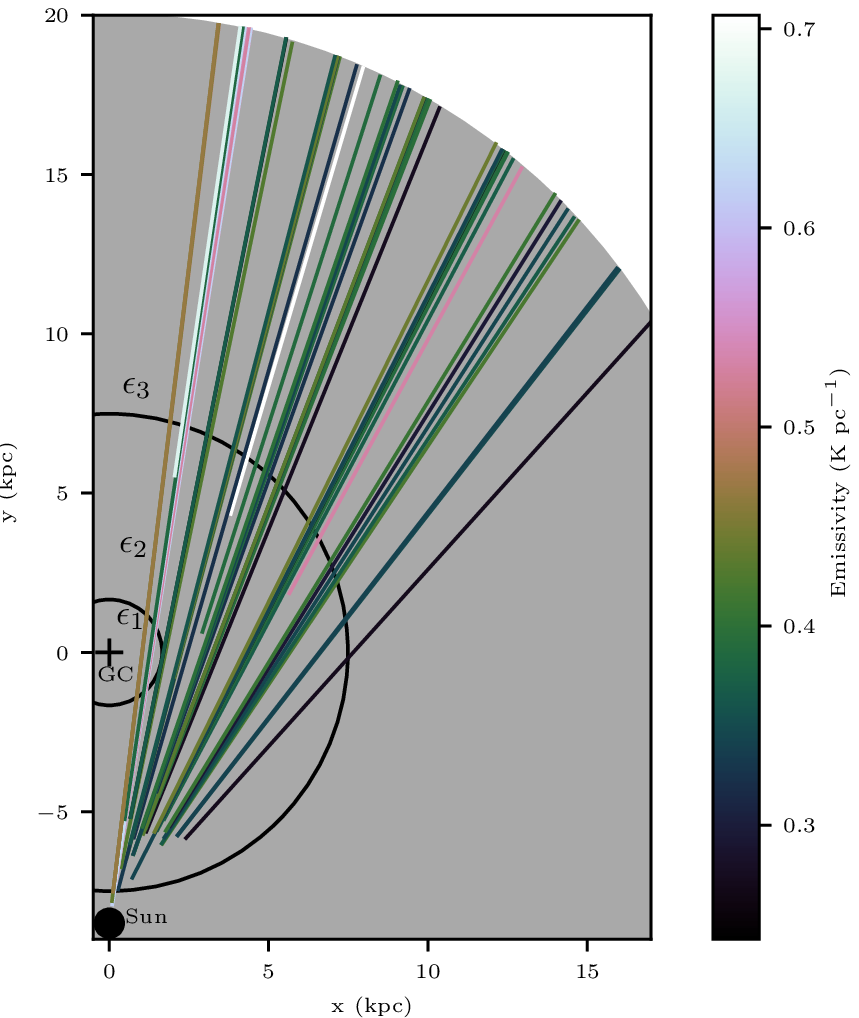}\hfill
\caption{Distribution of emissivities ($\epsilon_b$) on lines of sight and the Two-circle model. Each colour represents a path over which the emissivities are averaged. The colours of the lines indicate the value of emissivities along different paths. The Galactic non-thermal emission is assumed to lie in a disk with a Galactocentric radius of 20$\,\,$kpc shown by the grey circle segment. The two circles near the Galactic centre divide the Galactic plane into three regions with emissivities of $\epsilon_1$, $\epsilon_2$, and $\epsilon_3$ respectively. The Galactocentric radius of the Sun is 8.5$\,\,$kpc. The Sun and Galactic centre (GC) are marked as a black dot and plus respectively.} 
\label{fig:cir2} 
\end{figure*}

\begin{figure*}
\includegraphics[width=0.48\textwidth]{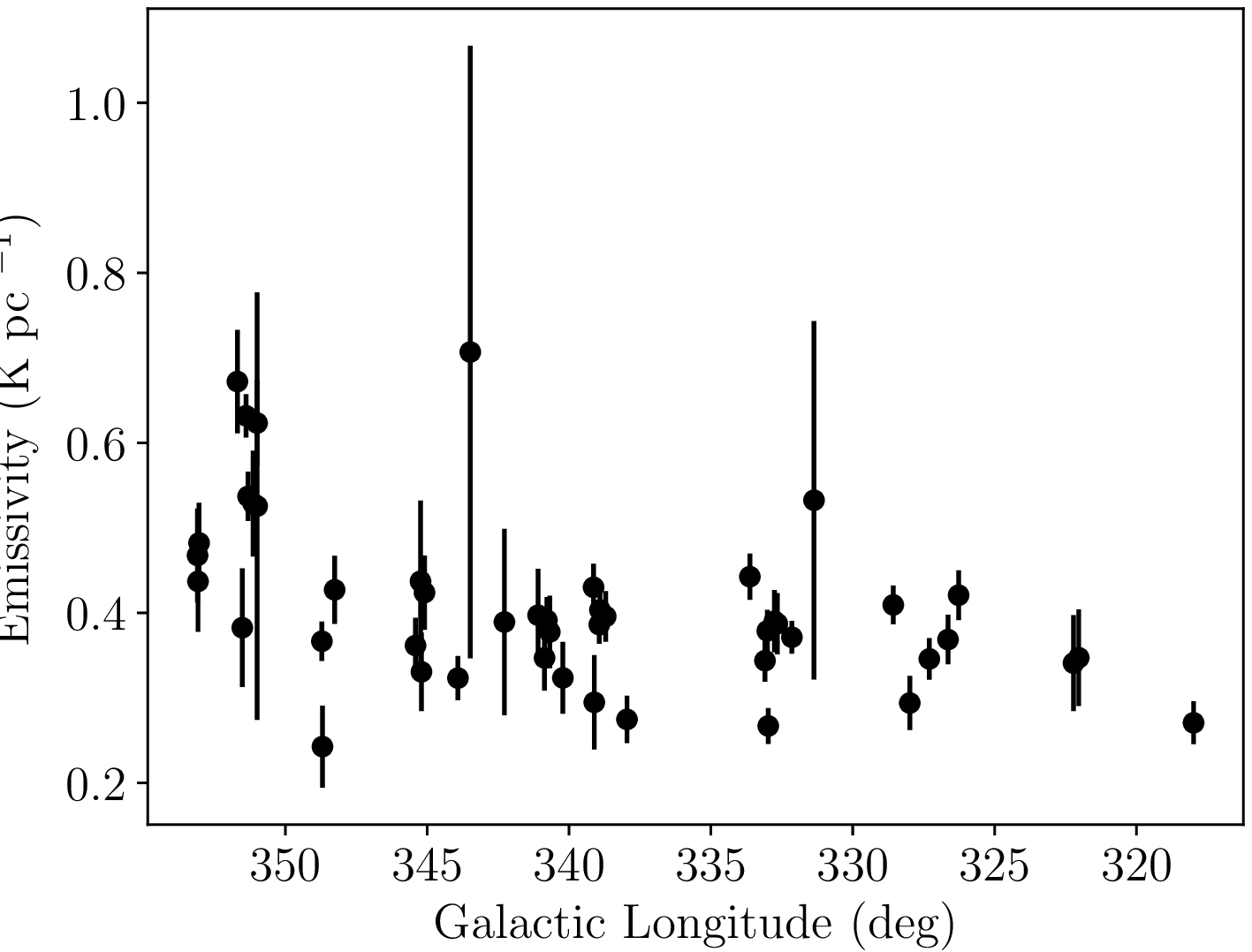}
\includegraphics[width=0.48\textwidth]{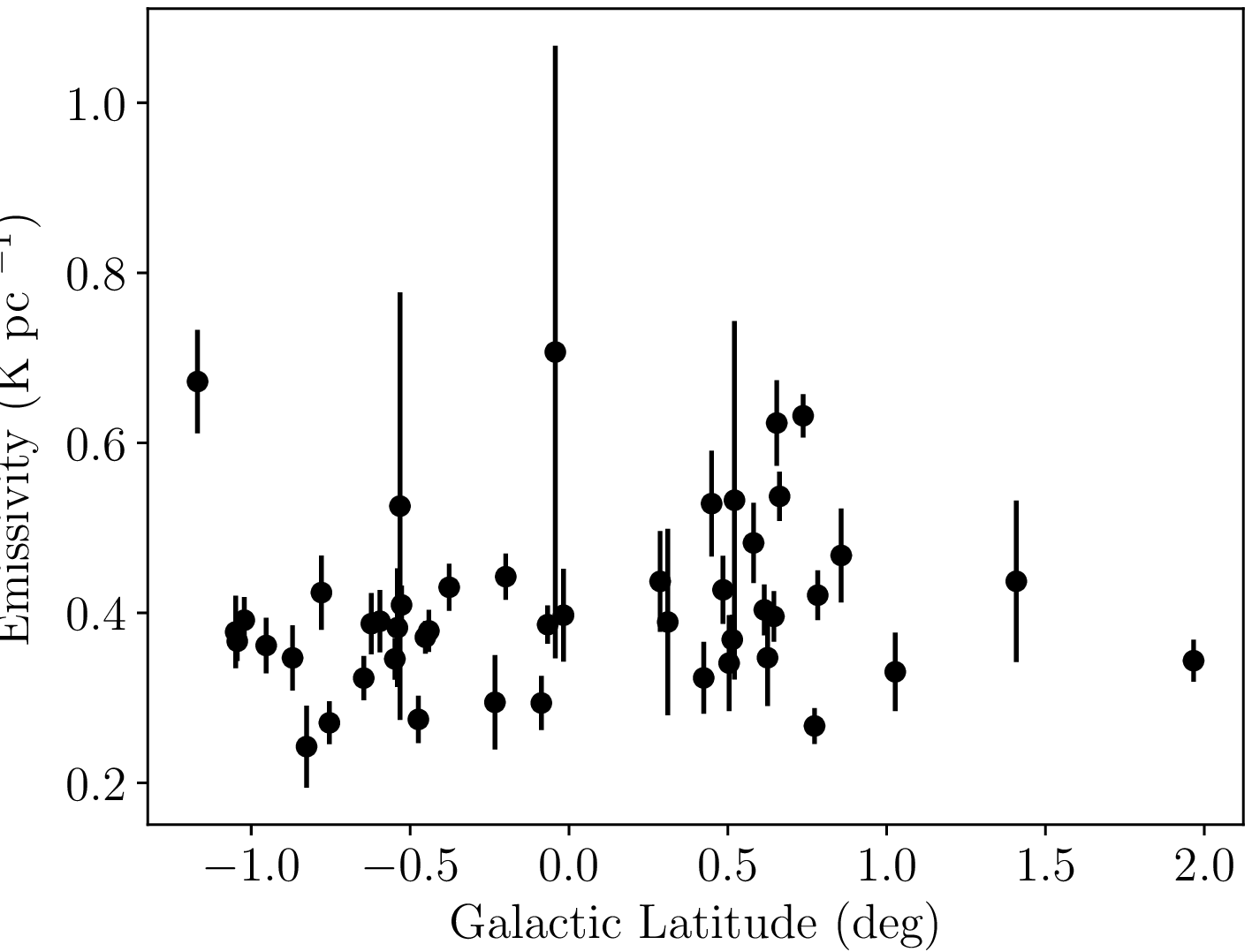}
\caption{Distributions of emissivities ($\epsilon_b$) with respect to Galactic longitude (left) and latitude (right). Some measurements have large errors due to the uncertainty of the distances of \hii regions.}
\label{fig:emi_dis_gl} 
\end{figure*}

\begin{figure}
\includegraphics[width=0.48\textwidth]{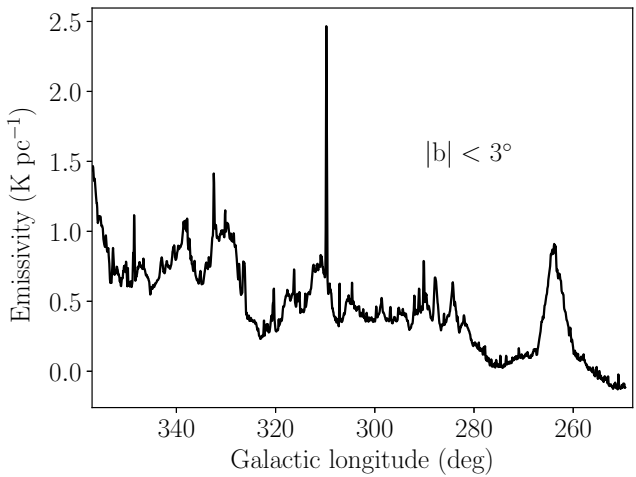}
\caption{Distributions of emissivity from the Sun to the Galactic edge over the line of sight with respect to the Galactic longitude in the range $250^\circ<l<355^\circ, |b|<3^\circ$: this includes the contributions from point sources and supernova remnants. The bin size in Galactic longitude is 3\farcm75. There is a trend of increasing emissivity with longitude. The high emissivity near the Galactic longitude 264$^\circ$ is contributed by the Vela supernova remnant. Note that we do not recover the total power of the Galaxy due to the interferometric nature of the measurements.}
\label{fig:distr_total_l_K} 
\end{figure}

 Fig.~\ref{fig:cir2} shows the emissivity ($\epsilon_b$) distribution along different lines of sight. Our measured emissivities increase with respect to Galactic longitude (Fig.~\ref{fig:emi_dis_gl}, left). This is coincident with the total line-of-sight emissivity distribution, which can be seen in Fig.~\ref{fig:distr_total_l_K}. The emissivity over the line of sight excluding point sources and supernova remnants decreases obviously with respect to Galactic longitude from 355$^\circ$ to 250$^\circ$. Our measured emissivities are relatively high measured from \hii regions with high Galactic latitude and relatively low derived from \hii regions with low Galactic latitude (Fig.~\ref{fig:emi_dis_gl}, right). This plot possibly reflects the decreasing of emissivity with the distance to the Galactic plane \citep{Peterson2002ApJ...575..217P}, although our detections are limited in the range of the Galactic latitude from -1$^\circ$ to 2$^\circ$.   
 
 Similar variations are also found in other surveys, e.g. the 408$\,\,$MHz Haslam map (Fig. 4 in \citealt{Haslam1982A&AS...47....1H}). The intensity profiles along Galactic longitude and latitude are well fitted by three magnetic field models in \citet{Sun2008A&A...477..573S} and \citet{Sun2010RAA....10.1287S}.
 
\subsection{Comparison with the literature} 
According to the brightness temperature spectrum $T(\nu)~\varpropto~\nu^{-2.3}$ \citep{Guzman2011A&A...525A.138G} and the typical emissivity of 0.01$\,\,$K$\,\,$pc$^{-1}$ \citep{Beuermann1985A&A...153...17B} at 408$\,\,$MHz \citep{Haslam1982A&AS...47....1H}, the typical emissivity at 88$\,\,$MHz is expected to be 0.34$\,\,$K$\,\,$pc$^{-1}$. The mean of our measured emissivity ($\epsilon_b$) of 0.40$\,\,$K$\,\,$pc$^{-1}$ with a variance of 0.10$\,\,$K$\,\,$pc$^{-1}$ agrees within 1$\sigma$ with this estimation. 

To check for any systematic differences between the samples, we compare the measurements for two \hii regions that appear both in our sample and N06, in Table~\ref{tab:compare}. Using the same distances and electron temperatures as listed in N06, and scaling our measurements from 88 to 74~MHz using a brightness temperature spectral index of 2.3 ($T_\nu~\varpropto~\nu^{-2.3}$, \citealt{Guzman2011A&A...525A.138G}), our measurements agree within 1.5$\sigma$ with those in N06.

\begin{table*}
\centering
\begin{threeparttable}
\caption{Emissivity measurements for \hii regions with absorption features.} 
\begin{tabular}{cccccrlrlccc}
\hline
WISE name & $I_o$ & $I_{t}$ & $T_o$ & $T_t$ & $D_f\,\,\,\,\,$ & $\,\,\,\,\,\,\,\,\,\,\,\,T_e$ & $\,\,\epsilon_f\,\,\,\,\,\,\,\,\,$ & $\,\,\,\,\,\,\,\,\,\epsilon_b$ & Q & Ref. \tabularnewline
          & Jy beam$^{-1}$ & Jy beam$^{-1}$ & $\times$10$^4$ K & $\times$10$^4$ K & kpc$\,\,\,\,\,$ & $\,\,\,\,\,\,\times$10$^3$ K & $\,\,$K pc$^{-1}\,\,$ & $\,\,\,\,$K pc$^{-1}$ &  & \tabularnewline
(1) & (2) & (3) & (4) & (5) & (6)$\,\,\,\,\,\,$ & $\,\,\,\,\,\,\,\,\,\,\,\,$(7) & $\,\,\,\,\,\,$(8)$\,\,\,\,\,\,\,\,\,$ & $\,\,\,\,\,\,\,\,\,$(9) & (10) & (11)\tabularnewline
\hline
G317.988$-$00.754 & 1.86$\pm$0.14 & 2.80$\pm$0.14 & 1.05$\pm$0.08 & 1.58$\pm$0.08 & 3.6$\pm$1.1 & 4.60$\pm$0.37 & -0.55$\pm$0.20 & 0.27$\pm$0.03 & 6 & 5;5\tabularnewline
G322.036$+$00.625 & 1.30$\pm$0.11 & 1.67$\pm$0.11 & 0.73$\pm$0.06 & 0.94$\pm$0.06 & 3.5$\pm$3.5 & 7.29$\pm$0.33$^{a}$ & -1.56$\pm$1.56 & 0.35$\pm$0.06 & 2 & 1;2\tabularnewline
G322.220$+$00.504 & 1.45$\pm$0.14 & 1.72$\pm$0.14 & 0.82$\pm$0.08 & 0.97$\pm$0.08 & 3.5$\pm$3.5 & 7.29$\pm$0.33$^{a}$ & -1.50$\pm$1.50 & 0.34$\pm$0.06 & 1 & 1;2\tabularnewline
G326.270$+$00.783 & 1.49$\pm$0.29 & 3.30$\pm$0.29 & 0.84$\pm$0.16 & 1.87$\pm$0.16 & 3.0$\pm$0.4 & 7.33$\pm$0.33$^{a}$ & -1.74$\pm$0.29 & 0.42$\pm$0.03 & 5 & 1;2\tabularnewline
G326.643$+$00.514 & 2.35$\pm$0.29 & 3.30$\pm$0.29 & 1.33$\pm$0.16 & 1.87$\pm$0.16 & 3.0$\pm$0.4 & 7.32$\pm$0.33$^{a}$ & -1.33$\pm$0.25 & 0.37$\pm$0.03 & 4 & 1;2\tabularnewline
G327.300$-$00.548 & 1.32$\pm$0.21 & 2.73$\pm$0.21 & 0.75$\pm$0.12 & 1.54$\pm$0.12 & 3.2$\pm$0.4 & 6.10$\pm$0.36 & -1.32$\pm$0.22 & 0.35$\pm$0.02 & 6 & 1,7;5\tabularnewline
G327.991$-$00.087 & 5.12$\pm$0.18 & 5.68$\pm$0.18 & 2.89$\pm$0.10 & 3.21$\pm$0.10 & 3.6$\pm$1.8 & 6.00$\pm$0.36 & 0.34$\pm$0.21 & 0.29$\pm$0.03 & 5 & 5;5\tabularnewline
G328.572$-$00.527 & 4.29$\pm$0.19 & 5.98$\pm$0.19 & 2.42$\pm$0.11 & 3.38$\pm$0.11 & 3.4$\pm$0.4 & 7.19$\pm$0.33$^{a}$ & -0.33$\pm$0.13 & 0.41$\pm$0.02 & 4 & 1;2\tabularnewline
G331.365$+$00.521 & 3.32$\pm$0.26 & 5.65$\pm$0.26 & 1.88$\pm$0.15 & 3.19$\pm$0.15 & 11.8$\pm$5.9 & 4.80$\pm$0.34 & -0.01$\pm$0.04 & 0.53$\pm$0.21 & 6 & 5;5\tabularnewline
G332.145$-$00.452 & 3.44$\pm$0.14 & 4.60$\pm$0.14 & 1.94$\pm$0.08 & 2.60$\pm$0.08 & 3.7$\pm$0.4 & 7.05$\pm$0.32$^{a}$ & -0.59$\pm$0.12 & 0.37$\pm$0.02 & 4 & 1;2\tabularnewline
G332.657$-$00.622 & 2.19$\pm$0.39 & 3.68$\pm$0.39 & 1.24$\pm$0.22 & 2.08$\pm$0.22 & 3.3$\pm$0.4 & 7.15$\pm$0.32$^{a}$ & -1.23$\pm$0.24 & 0.39$\pm$0.04 & 3 & 1;2\tabularnewline
G332.762$-$00.595 & 2.18$\pm$0.39 & 3.68$\pm$0.39 & 1.23$\pm$0.22 & 2.08$\pm$0.22 & 3.8$\pm$0.4 & 7.01$\pm$0.32$^{a}$ & -1.03$\pm$0.20 & 0.39$\pm$0.04 & 3 & 1;2\tabularnewline
G332.978$+$00.773 & 4.16$\pm$0.16 & 5.75$\pm$0.16 & 2.35$\pm$0.09 & 3.25$\pm$0.09 & 3.8$\pm$0.5 & 4.00$\pm$0.35 & 0.50$\pm$0.13 & 0.27$\pm$0.02 & 5 & 5;5\tabularnewline
G333.011$-$00.441 & 2.09$\pm$0.23 & 3.42$\pm$0.23 & 1.18$\pm$0.13 & 1.93$\pm$0.13 & 3.6$\pm$0.4 & 7.06$\pm$0.32$^{a}$ & -1.14$\pm$0.18 & 0.38$\pm$0.02 & 5 & 1;2\tabularnewline
G333.093$+$01.966 & 1.77$\pm$0.24 & 2.57$\pm$0.24 & 1.00$\pm$0.14 & 1.45$\pm$0.14 & 1.6$\pm$0.6 & 7.67$\pm$0.35$^{a}$ & -3.23$\pm$1.25 & 0.34$\pm$0.02 & 5 & 1;2\tabularnewline
G333.627$-$00.199 & 2.42$\pm$0.27 & 4.90$\pm$0.27 & 1.37$\pm$0.15 & 2.77$\pm$0.15 & 3.2$\pm$0.4 & 7.16$\pm$0.32$^{a}$ & -1.17$\pm$0.21 & 0.44$\pm$0.03 & 4 & 1;2\tabularnewline
G337.957$-$00.474 & 4.66$\pm$0.19 & 5.46$\pm$0.19 & 2.63$\pm$0.11 & 3.09$\pm$0.11 & 3.1$\pm$1.6 & 5.60$\pm$0.35 & 0.32$\pm$0.22 & 0.27$\pm$0.03 & 4 & 5;5\tabularnewline
G338.706$+$00.645 & 3.87$\pm$0.30 & 5.64$\pm$0.30 & 2.19$\pm$0.17 & 3.19$\pm$0.17 & 4.3$\pm$0.4 & 6.76$\pm$0.31$^{a}$ & -0.30$\pm$0.13 & 0.40$\pm$0.03 & 4 & 1;2\tabularnewline
G338.911$+$00.615 & 3.75$\pm$0.30 & 5.64$\pm$0.30 & 2.12$\pm$0.17 & 3.19$\pm$0.17 & 4.4$\pm$0.4 & 6.73$\pm$0.31$^{a}$ & -0.32$\pm$0.12 & 0.40$\pm$0.03 & 4 & 1;2\tabularnewline
G338.934$-$00.067 & 4.62$\pm$0.21 & 6.29$\pm$0.21 & 2.61$\pm$0.12 & 3.56$\pm$0.12 & 3.2$\pm$0.4 & 7.10$\pm$0.32$^{a}$ & -0.18$\pm$0.14 & 0.39$\pm$0.02 & 4 & 1;2\tabularnewline
G339.109$-$00.233 & 4.25$\pm$0.29 & 5.70$\pm$0.29 & 2.40$\pm$0.16 & 3.22$\pm$0.16 & 6.5$\pm$3.3 & 4.20$\pm$0.32 & 0.28$\pm$0.16 & 0.29$\pm$0.06 & 4 & 5;5\tabularnewline
G339.134$-$00.377 & 3.25$\pm$0.29 & 5.70$\pm$0.29 & 1.84$\pm$0.16 & 3.22$\pm$0.16 & 3.0$\pm$0.4 & 7.16$\pm$0.32$^{a}$ & -0.86$\pm$0.21 & 0.43$\pm$0.03 & 3 & 1;2\tabularnewline
G340.216$+$00.424 & 2.75$\pm$0.30 & 4.71$\pm$0.30 & 1.55$\pm$0.17 & 2.66$\pm$0.17 & 4.4$\pm$2.2 & 4.80$\pm$0.33 & -0.21$\pm$0.16 & 0.32$\pm$0.04 & 6 & 5;5\tabularnewline
G340.678$-$01.049 & 2.22$\pm$0.28 & 3.81$\pm$0.28 & 1.25$\pm$0.16 & 2.15$\pm$0.16 & 2.3$\pm$2.3 & 7.38$\pm$0.33$^{a}$ & -1.84$\pm$1.86 & 0.38$\pm$0.04 & 1 & 1;2\tabularnewline
G340.780$-$01.022 & 1.89$\pm$0.28 & 3.82$\pm$0.28 & 1.12$\pm$0.16 & 2.16$\pm$0.16 & 2.3$\pm$0.6 & 7.38$\pm$0.33$^{a}$ & -1.99$\pm$0.57 & 0.39$\pm$0.03 & 2 & 1;2\tabularnewline
G340.862$-$00.870 & 3.22$\pm$0.23 & 4.26$\pm$0.23 & 1.82$\pm$0.13 & 2.41$\pm$0.13 & 2.3$\pm$2.3 & 7.38$\pm$0.33$^{a}$ & -1.23$\pm$1.25 & 0.35$\pm$0.04 & 1 & 1;2\tabularnewline
G341.090$-$00.017 & 3.22$\pm$0.15 & 5.16$\pm$0.15 & 1.82$\pm$0.08 & 2.92$\pm$0.08 & 3.2$\pm$3.2 & 7.07$\pm$0.32$^{a}$ & -0.79$\pm$0.80 & 0.40$\pm$0.05 & 3 & 1;2\tabularnewline
G342.277$+$00.311 & 2.97$\pm$0.33 & 5.25$\pm$0.33 & 1.68$\pm$0.19 & 2.97$\pm$0.19 & 9.6$\pm$4.8 & 3.90$\pm$0.32 & 0.03$\pm$0.06 & 0.39$\pm$0.11 & 4 & 5;5\tabularnewline
G343.480$-$00.043 & 2.46$\pm$0.21 & 4.03$\pm$0.21 & 1.39$\pm$0.12 & 2.28$\pm$0.12 & 13.4$\pm$7.4 & 8.10$\pm$0.35 & -0.34$\pm$0.19 & 0.71$\pm$0.36 & 6 & 5;5\tabularnewline
G343.914$-$00.646 & 3.71$\pm$0.16 & 4.38$\pm$0.16 & 2.10$\pm$0.09 & 2.48$\pm$0.09 & 2.8$\pm$1.4 & 7.20$\pm$0.35 & -0.70$\pm$0.38 & 0.32$\pm$0.03 & 5 & 5;5\tabularnewline
G345.094$-$00.779 & 1.72$\pm$0.29 & 4.26$\pm$0.33 & 0.97$\pm$0.19 & 2.41$\pm$0.19 & 2.1$\pm$2.1 & 7.43$\pm$0.33$^{a}$ & -2.38$\pm$2.39 & 0.42$\pm$0.04 & 4 & 1;2\tabularnewline
G345.202$+$01.027 & 1.18$\pm$0.61 & 4.10$\pm$0.61 & 0.67$\pm$0.34 & 2.32$\pm$0.34 & 1.1$\pm$0.6 & 4.80$\pm$0.12 & -2.85$\pm$1.74 & 0.33$\pm$0.05 & 4 & 4;5\tabularnewline
G345.235$+$01.408 & 1.12$\pm$0.34 & 2.59$\pm$0.34 & 0.69$\pm$0.19 & 1.46$\pm$0.19 & 8.0$\pm$4.0 & 6.00$\pm$0.35 & -0.64$\pm$0.33 & 0.44$\pm$0.09 & 5 & 2;2\tabularnewline
G345.410$-$00.953 & 2.00$\pm$0.40 & 3.60$\pm$0.40 & 1.13$\pm$0.23 & 2.03$\pm$0.23 & 2.6$\pm$0.6 & 6.96$\pm$0.05 & -1.59$\pm$0.43 & 0.36$\pm$0.03 & 6 & 1;2\tabularnewline
G348.261$+$00.485 & 3.40$\pm$0.32 & 6.05$\pm$0.32 & 1.92$\pm$0.18 & 3.42$\pm$0.18 & 1.8$\pm$1.8 & 7.53$\pm$0.34$^{a}$ & -1.51$\pm$1.55 & 0.43$\pm$0.04 & 4 & 1;2\tabularnewline
G348.691$-$00.826 & 2.15$\pm$0.33 & 3.03$\pm$0.33 & 1.22$\pm$0.19 & 1.71$\pm$0.14 & 3.4$\pm$0.3 & 4.80$\pm$1.00 & -0.52$\pm$0.33 & 0.24$\pm$0.05 & 6 & 1;6\tabularnewline
G348.710$-$01.044 & 1.27$\pm$0.28 & 2.67$\pm$0.28 & 0.72$\pm$0.16 & 1.51$\pm$0.16 & 3.4$\pm$0.3 & 6.20$\pm$1.00 & -1.57$\pm$0.18 & 0.37$\pm$0.02 & 6 & 1;8\tabularnewline
G350.991$-$00.532 & 3.16$\pm$0.31 & 4.31$\pm$0.31 & 1.79$\pm$0.18 & 2.44$\pm$0.18 & 13.7$\pm$6.9 & 6.10$\pm$0.35 & -0.12$\pm$0.07 & 0.53$\pm$0.25 & 5 & 5;5\tabularnewline
G350.995$+$00.654 & 2.08$\pm$0.67 & 6.86$\pm$0.67 & 1.18$\pm$0.38 & 3.88$\pm$0.38 & 0.6$\pm$0.3 & 10.57$\pm$0.34 & -12.72$\pm$6.58 & 0.62$\pm$0.05 & 5 & 8;8\tabularnewline
G351.130$+$00.449 & 2.85$\pm$0.82 & 8.24$\pm$0.82 & 1.61$\pm$0.46 & 4.66$\pm$0.46 & 1.4$\pm$0.7 & 6.65$\pm$0.07 & -1.87$\pm$1.25 & 0.53$\pm$0.06 & 5 & 8;8\tabularnewline
G351.311$+$00.663 & 2.12$\pm$0.35 & 6.96$\pm$0.35 & 1.20$\pm$0.20 & 3.93$\pm$0.20 & 1.3$\pm$0.1 & 7.71$\pm$0.35$^{a}$ & -3.63$\pm$0.54 & 0.54$\pm$0.03 & 3 & 9;2\tabularnewline
G351.383$+$00.737 & 1.77$\pm$0.34 & 7.02$\pm$0.34 & 1.00$\pm$0.19 & 3.97$\pm$0.19 & 1.3$\pm$0.1 & 9.70$\pm$0.09 & -5.54$\pm$0.57 & 0.63$\pm$0.03 & 5 & 9;2\tabularnewline
G351.516$-$00.540 & 3.28$\pm$0.30 & 6.04$\pm$0.38 & 1.85$\pm$0.17 & 3.41$\pm$0.21 & 3.3$\pm$3.3 & 5.70$\pm$1.00 & -0.32$\pm$0.46 & 0.38$\pm$0.07 & 3 & 6;3\tabularnewline
G351.688$-$01.169 & 1.71$\pm$0.25 & 3.87$\pm$0.25 & 0.97$\pm$0.14 & 2.19$\pm$0.14 & 14.2$\pm$1.0 & 6.49$\pm$0.21 & -0.29$\pm$0.04 & 0.67$\pm$0.06 & 6 & 2;2\tabularnewline
G353.038$+$00.581 & 1.52$\pm$0.56 & 5.33$\pm$0.56 & 0.86$\pm$0.32 & 3.01$\pm$0.32 & 1.1$\pm$1.1 & 7.78$\pm$0.35$^{a}$ & -5.12$\pm$5.18 & 0.48$\pm$0.05 & 3 & 1;2\tabularnewline
G353.076$+$00.287 & 2.06$\pm$0.75 & 6.81$\pm$0.75 & 1.16$\pm$0.42 & 3.85$\pm$0.42 & 0.7$\pm$1.5 & 5.39$\pm$0.10 & -3.54$\pm$7.74 & 0.44$\pm$0.06 & 5 & 2;2\tabularnewline
G353.092$+$00.857 & 1.29$\pm$0.56 & 5.33$\pm$0.56 & 0.73$\pm$0.32 & 3.01$\pm$0.32 & 1.0$\pm$2.0 & 7.10$\pm$0.40 & -5.28$\pm$10.59 & 0.47$\pm$0.06 & 5 & 2;2\tabularnewline
\hline
\end{tabular}
\begin{tablenotes}
\item Notes: Col. (1): The name of \hii regions from the WISE \hii region catalogue. Cols. (2) and (3): The MWA surface brightness measurements of the absorbing \hii region and the neighbouring region, respectively. Cols. (4) and (5): The brightness temperatures of the absorbing \hii region and the neighbouring region, respectively. Col. (6): The distance from \hii regions to the Sun found in the literature. Col. (7): The electron temperature of \hii regions from the literature. Col. (8): The average emissivity from \hii regions to the Sun. 
Col. (9): The average emissivity from \hii regions to the Galactic edge. Col. (10): The indicative quality of emissivity measurements: a higher value means higher quality (see Section~\ref{sec:emi_distri}). Col. (11): The references for the distances and electron temperatures. The first and second number indicate the reference for the distance in Col. (6) and the electron temperature in Col. (7) respectively. The reference list: 
1. \citet{Anderson2014ApJS..212....1A}; 
2. \citet{Balser2015ApJ...806..199B}; 
3. \citet{Caswell1987A&A...171..261C}; 
4. \citet{Garcia2014ApJS..212....2G}; 
5. \citet{Hou2014A&A...569A.125H}; 
6. \citet{Nord2006AJ....132..242N}; 
7. \citet{Paladini2004MNRAS.347..237P}; 
8. \citet{Quireza2006ApJ...653.1226Q}; 
9. \citet{Reid2014ApJ...783..130R}.

$^a$ The electron temperature of this \hii region is derived from the statistical relation $T_e\,\,=\,\,$(4928$\,\,\pm\,\,$277)$\,\,+\,\,$(385$\,\,\pm\,\,$29)$\,\,R_{\textrm{gal}}$ from \citet{Balser2015ApJ...806..199B}.
\end{tablenotes}
\label{tab:emi}
\end{threeparttable}
\end{table*}

\begin{table*}
\begin{threeparttable}
%\captionsetup{justification=centering}
\centering
\caption{Comparison of two emissivities measured in N06 and this work.}
\begin{tabular}{ccccc}
\hline 
\hii region (this work) & G351.516$-$00.540 & G348.710$-$01.044 & G348.691$-$00.826 \tabularnewline
\hii region (N06) & G351.5$-$0.5 & G348.7$-$1.0 & G348.6$-$0.6\tabularnewline
Distance (kpc, used in N06) & 3.3 & 2.0 & 2.7\tabularnewline
T$_e$ ($\times$10$^{3}\,\,$K, used in N06)& 5.7 $\pm$ 1.0 & 6.2 $\pm$ 1.0 & 4.8 $\pm$ 1.0\tabularnewline
Angular diameter (arcmin, used in N06) & 3.46 & 4.47 & 13.27\tabularnewline 
$\epsilon_{b}$(K pc$^{-1}$, at 74$\,\,$MHz, N06) & 0.36 $\pm$ 0.06 & 0.36 $\pm$ 0.05 & 0.51 $\pm$ 0.05\tabularnewline
$\epsilon_{b}$(K pc$^{-1}$, scaled at 74$\,\,$MHz, this work)\tnote{1} & 0.45 $\pm$ 0.06 & 0.35 $\pm$ 0.05 & 0.26 $\pm$ 0.05\tabularnewline
\hline 
\end{tabular}
\label{tab:compare}
\begin{tablenotes}
\item[1] To compare our measurements with those in N06, we re-calculate our emissivities for the two \hii regions using the distances and electron temperatures from N06 and scale our measurements from 88 to 74 MHz using a brightness temperature spectral index of 2.3. 
\end{tablenotes}
\end{threeparttable}
\end{table*}

\subsection{Non-detections and Detection Bias}
 About 80~\hii regions with radii larger than the beam size and smaller than the maximum angular scale sensitivity are not detected as absorption regions in the region we analyse (Fig.~\ref{fig:GP_img}). A possible reason is that not enough synchrotron emission exists behind these \hii regions; an absorption feature is only apparent when the integrated brightness temperature blocked by an \hii region is larger than its electron temperature (see Equations~\ref{eq:To} and \ref{eq:Tt}). As show in Fig.~\ref{fig:GP_img}, all of our 47~detections are in the range of $315^\circ<l<355^\circ$ whereas only 25~non-detections are in this range. The other 55~non-detections are in the low brightness temperature region with the range of $250^\circ<l<315^\circ$, which might be detectable in emission at higher frequencies and these will be followed up in future. Two other possibilities of non-detections are a high electron temperature and/or a low optical depth for these \hii regions.

Note that we assume \hii regions associated with absorption features have an optical depth much larger than one. In future, the work will use the MWA GLEAM survey coverage 72--231$\,\,$MHz to measure the optical depth and confirm this hypothesis.
% In the case of optical depth close to one, the emissivity $\epsilon_b$ becomes the lower limit and $\epsilon_f$ becomes the upper limit.
 
Finally, our emissivity measurements do not sample the whole Galactic disk uniformly mainly due to the non-uniform distribution of \hii regions. Most of the observed \hii regions are nearby, with distances of several kiloparsecs. The \hii regions with large distances usually have large distance uncertainties, resulting in large emissivity uncertainties.       

\section{Modelling the emissivity distribution}
\label{sec:modelling}
The measured emissivities need modelling to show the emissivity distribution with respect to the Galactic radius. We build four simple models to probe this complex problem.  

The synchrotron emissivity is the power per unit volume per unit frequency per unit solid angle produced by cosmic-ray electrons interacting with the magnetic field. Since we can only measure the average emissivity along paths, we adopt four simple models: Uniform; Gaussian; Exponential; and Two-circle to fit the emissivity distribution of the Galaxy. We fit our measured $\epsilon_b$ using these models. We test a Uniform model given that N06 found a constant emissivity outside of a circular region centred at the Galactic centre with a radius of 3 kpc. The Exponential model is motivated by the exponential cosmic-ray electron distribution used in \citet{Sun2008A&A...477..573S}. The Gaussian model is an alternative form to express a decreasing emissivity profile from the Galactic centre. The two-circle model is constructed with more free parameters and to specifically account for the high emissivities found for regions with paths close to the Galactic centre.

 Before showing details of each model, we define $R_0$ to be the Galactocentric radius of the Sun, $R$ to be the Galactocentric radius of any point on the Galactic plane, $d$ to be the distance from an \hii region to the Sun, $l$ to be the Galactic longitude, $\alpha$ and $\beta$ to be free parameters related to Gaussian and Exponential models. 

 The Uniform model assumes a constant emissivity distribution across the Galactic disk. 

 In the Gaussian model, the emissivity is assumed to have a distribution $\epsilon(R) = \alpha_1 \exp(-R^2/2\beta_1^2)$. We assume the emissivity peaks at the Galactic centre. $R$ is a function of the distance $d$, so the emissivity~$\epsilon$ is also a function of $d$. The average emissivity from the \hii region to the Galactic edge is the integration of $\epsilon(d)$ for $d$ and then divided by the corresponding path length. 

 The Exponential model has an emissivity distribution in the form of a natural exponential function $\epsilon(R)~=~\alpha_2\exp(-\beta_2R)$. The method of calculating the average emissivities along the line of sight from this model is the same as that in the Gaussian model. 

We also develop a simple Two-circle model motivated by the data. In this model, we divide the Galactic plane into three regions using two circles centring on the Galactic centre. $R_1$ and $R_2$ ($R_1 < R_2$) is the radius of circle 1 and circle 2, respectively. $\epsilon_1$ is the emissivity in circle 1. $\epsilon_2$ is the emissivity in the region between circle 1 and circle 2. $\epsilon_3$ is the emissivity in the region outside circle 2 and within the Galactic edge (see Fig~\ref{fig:cir2}).    

 With the assumptions in each model, we determine the free parameters by fitting the observed emissivities ($\epsilon_b$) and those from the model using the (reduced) chi-squared test (e.g. \citealt{Andrae2010arXiv1012.3754A}) to compare the goodness of fit of the models. We did not fit the average emissivities from \hii regions to the Sun since they are not absolutely reliable as mentioned in Section~\ref{sec:data}. We sample the whole parameter spaces and find the minimum of chi-squares for each model. We use the boundary of fixed $\Delta\chi^2$ (2$\sigma$) to estimate the quoted errors. We project the full M-dimensional (M = 2, 5) confidence region on the one-dimensional confidence interval for each parameter.

\begin{figure*}
\centering
\includegraphics[width=\textwidth]{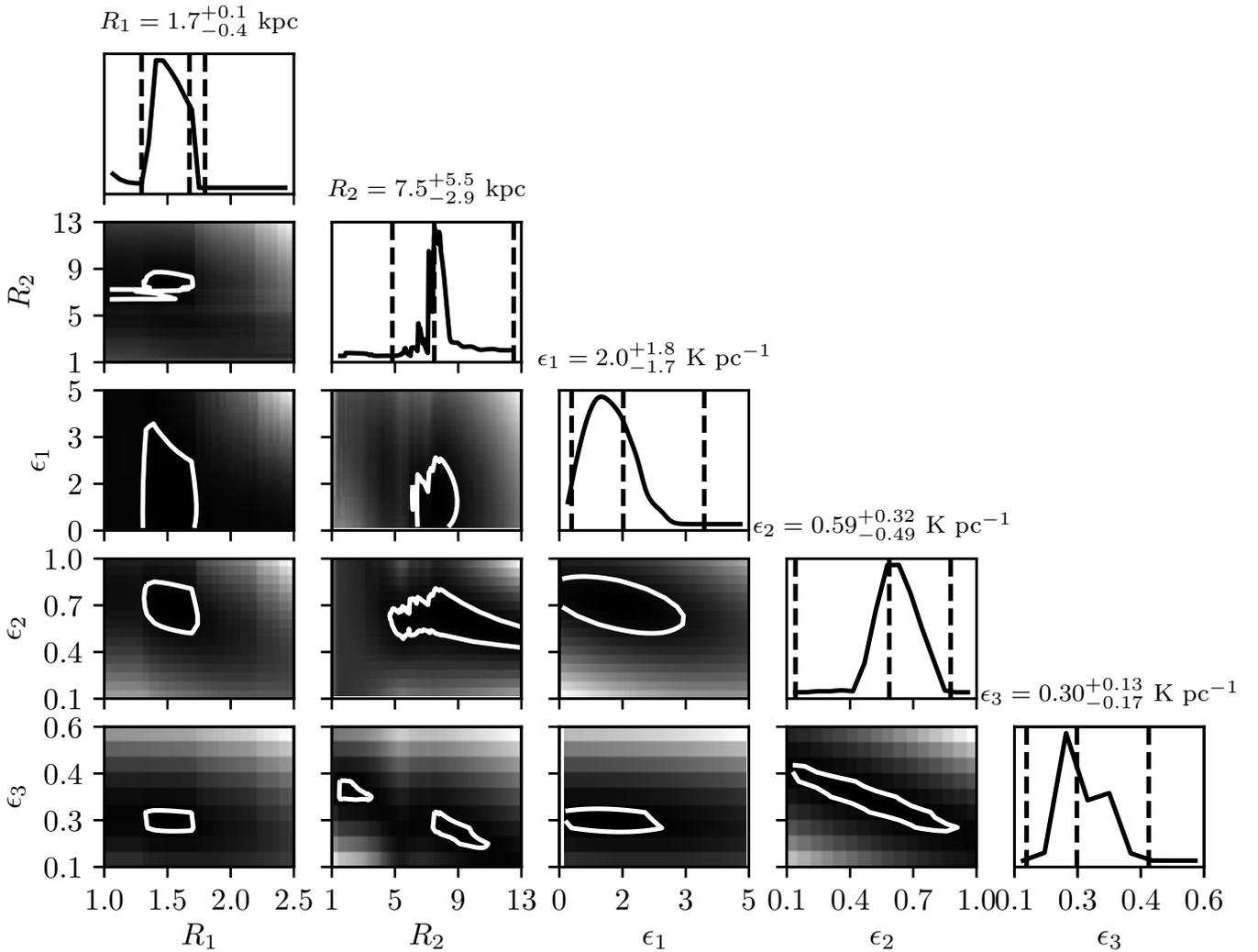}
\caption{Triangle plot showing the correlations between the two-circle model parameters: the grey scale traces the distributions of chi-square with darker regions indicating lower values. The 2$\sigma$ boundaries of chi-square are shown by white contours. The black solid lines show the relative likelihood distributions with dashed lines showing the best values and 2$\sigma$ limits of the parameters.}
\label{fig:model_triangle_5p}
\end{figure*}

 The results of modelling are shown in Table~\ref{tab:modelling}. The Uniform model shows an average emissivity of $0.38^{+0.04}_{-0.02}$ K pc$^{-1}$ with a poor reduced chi-square ($\chi^2_{\text{red}} = 6.51$). The Gaussian and Exponential models have similarly poor fits ($\chi^2_{\text{red}}\sim$ 6.03 and 5.77). The Two-circle model has the best performance compared with the other three models ($\chi^2_{\text{red}}=3.74$). It shows a high emissivity (2.0$^{+1.8}_{-1.7}\,\,$K$\,\,$pc$^{-1}$) region near the Galactic centre within a radius of 1.7$^{+0.1}_{-0.4}\,\,$kpc, a low emissivity (0.30$^{+0.13}_{-0.17}\,\,$K$\,\,$pc$^{-1}$) region outside of a radius of 7.5$^{+5.5}_{-2.9}\,\,$kpc, a medium emissivity (0.59$^{+0.32}_{-0.49}\,\,$K$\,\,$pc$^{-1}$) region between the two. In Fig.~\ref{fig:model_triangle_5p} we show the correlations between the parameters for the Two-Circle model. We see that in the 2$\sigma$ contour of chi-square, a lower $R_1$ corresponds to a higher $\epsilon_1$, but when $R_1$ is lower than about 1.3$\,\,$kpc, $\epsilon_1$ does not increase further. Similarly, a higher $R_2$ cannot give a lower $\epsilon_2$ when $R_2$ exceeds about 10$\,\,$kpc. In the sub-plot $R_2$:$\epsilon_3$, $R_2$ is fixed to be higher than 1$\,\,$kpc since it cannot be less than $R_1$. The 2$\sigma$ errors for each parameter in the Two-circle model are large due to the correlations between the parameters, and fundamentally due to the limited number of measurements, and because most of the \hii regions are nearby; those that are more distant have large error bars on their distance measurements, reducing their importance during the fitting process. Our best fitted $R_1$ of the innermost circle includes the brightest Galactic centre region. Potentially showing a physical correlation, the region between the best fitted $R_1$ and $R_2$ includes several spiral arms, e.g., the 3-kpc ring, the Sagittarius arm, and the Norma arm. The large region outside of the outer most circle only contains the Scutum-Centaurus arm and the McClure-Griffiths arm \citep{McClure-Griffiths2004ApJ...607L.127M}.       
 
\begin{figure*}
\centering
\includegraphics[width=\textwidth]{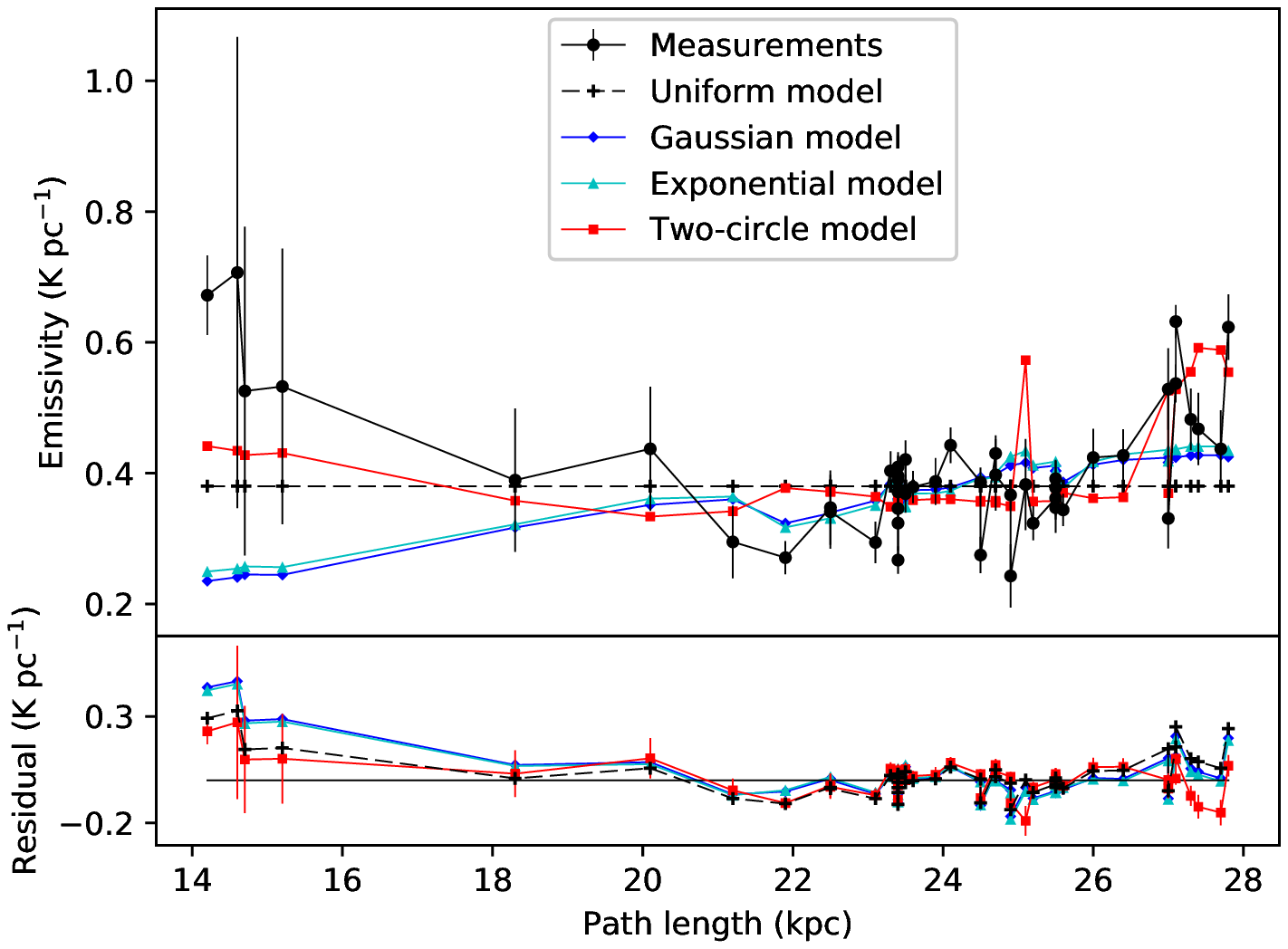}
\caption{Distributions of emissivities ($\epsilon_b$) from our measurements and models with the path length from \hii regions to the Galactic edge along the line of sight.}
\label{fig:models}
\end{figure*}

 Fig.~\ref{fig:models} compares the models in more detail where it can be seen that none provide a good fit to the four measurements with short path lengths (14$-$16$\,\,$kpc). However, all models similarly account for the measurements with longer path lengths (18$-$26$\,\,$kpc). The main differences among models appear at the path length above 26$\,\,$kpc, where there are several high average emissivities. These high emissivities pass the region near the Galactic centre. Only the Two-circle model, which has additional free parameters, can recover these high emissivities (2.0$^{+1.8}_{-1.7}\,\,$K$\,\,$pc$^{-1}$) near the Galactic centre.

\begin{table*}
\centering
\begin{threeparttable}
%\scriptsize %ref: https://en.wikibooks.org/wiki/LaTeX/Fonts
\caption{Parameters from fitting the measured emissivities with four models: Uniform, Gaussian, Exponential, and Two-circle model.}
\begin{tabular}{cccc}
\hline 
Model & Free parameters & $\chi_{\text{red}}^{2}$ & Degrees of freedom \tabularnewline
\hline
Uniform & $\epsilon=0.38^{+0.04}_{-0.02}$ & 6.51 & 46\tabularnewline
Gaussian & $\alpha_1=0.71^{+0.29}_{-0.22}$, $\beta_1=8.2^{+5.8}_{-2.7}$ & 6.03 & 45\tabularnewline
Exponential & $\alpha_2=0.94^{+0.46}_{-0.43}$, $\beta_2=0.11^{+0.06}_{-0.07}$ & 5.77 & 45\tabularnewline
Two-circle & $R_1=1.7^{+0.1}_{-0.4}$, $R_2=7.5^{+5.5}_{-2.9}$ & 3.74  & 42\tabularnewline
& $\epsilon_1=2.0^{+1.8}_{-1.7}$, $\epsilon_2=0.59^{+0.32}_{-0.49}$, $\epsilon_3=0.30^{+0.13}_{-0.17}$ \tabularnewline

\hline 
\end{tabular}
\begin{tablenotes}
\item Note: The unit of $\epsilon$, $\epsilon_1$, $\epsilon_2$, $\epsilon_3$, $\alpha_1$, and $\alpha_2$ are K~pc$^{-1}$. The unit of $R_1$, $R_2$, and $\beta_1$ are kiloparsecs. The unit of $\beta_2$ is kpc$^{-1}$. All the quoted errors are at $2\sigma$ level. The Gaussian model tends to the Uniform model when $\beta_1$ tends to infinity. 
%[$\epsilon$] = [$\epsilon_1$] = [$\epsilon_2$] = [$\epsilon_3$] = [$\alpha_1$] = [$\alpha_2$] = K~pc$^{-1}$; [$R_1$] = [$R_2$] = [$\beta_1$] = kpc; [$\beta_2$] = kpc$^{-1}$. 
\end{tablenotes}
\label{tab:modelling}
\end{threeparttable}
\end{table*}

%\begin{figure*}
%\centering
%\includegraphics[width=\textwidth]{fig/model_triangle_3p.eps}
%\caption{Triangle plot showing the correlations between the three emissivities in the Two-circle model when $R_1$ and $R_2$ are fixed at the best fitted value.}
%\label{fig:triangle_3p}
%\end{figure*}

 Furthermore, the Two-circle model can explain the low emissivity along the total lines of sight towards Galactic longitude between 270$^\circ$ and 320$^\circ$ (Fig.~\ref{fig:distr_total_l_K}). The brightness temperature in these directions is significantly lower compared with that in other directions in Fig.~\ref{fig:GP_img}. 
 
 N06 determined a best-fit model with uniform emissivity between 3 and 20$\,\,$kpc and zero within 3$\,\,$kpc of the Galactic centre. However, this model does not fit our data better than any of our models. Our best-fit Two-circle model shows the highest emissivity near the Galactic centre, contrary to N06 but quite likely due to sampling bias: our work samples regions in the fourth quadrant whereas N06's sample is mainly in the first quadrant; two of their measurements overlap with ours. Their measured emissivities are lower when the paths are closer to the Galactic centre, whereas ours are opposite, driving our model fits to increase the emissivity near the Galactic centre. Note that both of these two observations probe 2--3$\,\,$kpc nearby the Galactic centre but no detection is closer to the Galactic centre. The modelling in \citet{Sun2008A&A...477..573S} suggested a high magnetic field strength and the free electron model in NE2001 \citep{Cordes2002astro.ph..7156C} suggested a high free electron density near 2-3$\,\,$kpc of the Galactic centre. These results support our conclusions.

 Note that our four simplified models are only initial tests for the complex emissivity distribution. Our models have limited power to fully describe our measurements, not to mention other surveys at different frequencies. To understand the Galactic physical structures, we need to analysis our measurements and other survey data coherently using comprehensive models developed by \citet{Sun2008A&A...477..573S, Sun2010RAA....10.1287S, Orlando2013MNRAS.436.2127O}, etc. 

\section{Summary and Future work}
\label{sec:sum}
 We measured emissivities from 47~\hii regions along the two paths behind and in front of \hii regions. These measurements show the radial dependence of emissivity distribution. The modelling of these measurements shows a high emissivity region near the Galactic centre and low emissivities near the Galactic edge.

 In future, data covering two-thirds of the Galactic plane will be accessible through the GLEAM survey and we expect to derive new measurements from about 200~\hii regions. We will take advantage of the data in all five frequency bands of the GLEAM survey to derive the emissivity spectrum. A 3-D emissivity map, combining these discrete measurements and the total line-of-sight emissivity distributions, will show more details of the Galactic structures with the help of existed comprehensive 3-D models. We may set a length scale on cosmic ray electron propagation by comparing the possible cosmic ray source distribution and emissivity distribution. Possibly, we can derive the cosmic ray electron distribution with an assumed Galactic magnetic field distribution, and vice versa. 

\section*{Acknowledgements}
We thank the referee for a number of useful suggestions which improved this paper. This work makes use of the Murchison Radio-astronomy Observatory, operated by CSIRO. We acknowledge the Wajarri Yamatji people as the traditional owners of the Observatory site. Support for the operation of the MWA is provided by the Australian Government (NCRIS), under a contract to Curtin University administered by Astronomy Australia Limited. We acknowledge the Pawsey Supercomputing Centre which is supported by the Western Australian and Australian Governments. HS thanks the support from the NSFC (11473038, 11273025). CAJ thanks the Department of Science, Office of Premier \& Cabinet, WA for their support through the Western Australian Fellowship Program.

%%%%%%%%%%%%%%%%%%%%%%%%%%%%%%%%%%%%%%%%%%%%%%%%%%


% \citep[e.g.][]{Author2012}

\bibliographystyle{mnras}
\bibliography{bibtex}

\begin{thebibliography}{}
\makeatletter
\relax
\def\mn@urlcharsother{\let\do\@makeother \do\$\do\&\do\#\do\^\do\_\do\%\do\~}
\def\mn@doi{\begingroup\mn@urlcharsother \@ifnextchar [ {\mn@doi@}
  {\mn@doi@[]}}
\def\mn@doi@[#1]#2{\def\@tempa{#1}\ifx\@tempa\@empty \href
  {http://dx.doi.org/#2} {doi:#2}\else \href {http://dx.doi.org/#2} {#1}\fi
  \endgroup}
\def\mn@eprint#1#2{\mn@eprint@#1:#2::\@nil}
\def\mn@eprint@arXiv#1{\href {http://arxiv.org/abs/#1} {{\tt arXiv:#1}}}
\def\mn@eprint@dblp#1{\href {http://dblp.uni-trier.de/rec/bibtex/#1.xml}
  {dblp:#1}}
\def\mn@eprint@#1:#2:#3:#4\@nil{\def\@tempa {#1}\def\@tempb {#2}\def\@tempc
  {#3}\ifx \@tempc \@empty \let \@tempc \@tempb \let \@tempb \@tempa \fi \ifx
  \@tempb \@empty \def\@tempb {arXiv}\fi \@ifundefined
  {mn@eprint@\@tempb}{\@tempb:\@tempc}{\expandafter \expandafter \csname
  mn@eprint@\@tempb\endcsname \expandafter{\@tempc}}}

\bibitem[\protect\citeauthoryear{{Abramenkov} \& {Krymkin}}{{Abramenkov} \&
  {Krymkin}}{1990}]{Abramenkov1990IAUS..140...49A}
{Abramenkov} E.~A.,  {Krymkin} V.~V.,  1990, in {Beck} R.,  {Wielebinski} R.,
  {Kronberg} P.~P.,  eds,  IAU Symposium Vol. 140, Galactic and Intergalactic
  Magnetic Fields. pp 49--52

\bibitem[\protect\citeauthoryear{{Alves}, {Davies}, {Dickinson}, {Calabretta},
  {Davis}  \& {Staveley-Smith}}{{Alves}
  et~al.}{2012}]{Alves2012MNRAS.422.2429A}
{Alves} M.~I.~R.,  {Davies} R.~D.,  {Dickinson} C.,  {Calabretta} M.,  {Davis}
  R.,   {Staveley-Smith} L.,  2012, \mn@doi [\mnras]
  {10.1111/j.1365-2966.2012.20796.x}, \href
  {http://adsabs.harvard.edu/abs/2012MNRAS.422.2429A} {422, 2429}

\bibitem[\protect\citeauthoryear{{Anderson}, {Bania}, {Balser}, {Cunningham},
  {Wenger}, {Johnstone}  \& {Armentrout}}{{Anderson}
  et~al.}{2014}]{Anderson2014ApJS..212....1A}
{Anderson} L.~D.,  {Bania} T.~M.,  {Balser} D.~S.,  {Cunningham} V.,  {Wenger}
  T.~V.,  {Johnstone} B.~M.,   {Armentrout} W.~P.,  2014, \mn@doi [\apjs]
  {10.1088/0067-0049/212/1/1}, \href
  {http://adsabs.harvard.edu/abs/2014ApJS..212....1A} {212, 1}

\bibitem[\protect\citeauthoryear{{Andrae}, {Schulze-Hartung}  \&
  {Melchior}}{{Andrae} et~al.}{2010}]{Andrae2010arXiv1012.3754A}
{Andrae} R.,  {Schulze-Hartung} T.,   {Melchior} P.,  2010, preprint, \href
  {http://adsabs.harvard.edu/abs/2010arXiv1012.3754A} {} (\mn@eprint {arXiv}
  {1012.3754})

\bibitem[\protect\citeauthoryear{{Balser}, {Wenger}, {Anderson}  \&
  {Bania}}{{Balser} et~al.}{2015}]{Balser2015ApJ...806..199B}
{Balser} D.~S.,  {Wenger} T.~V.,  {Anderson} L.~D.,   {Bania} T.~M.,  2015,
  \mn@doi [\apj] {10.1088/0004-637X/806/2/199}, \href
  {http://adsabs.harvard.edu/abs/2015ApJ...806..199B} {806, 199}

\bibitem[\protect\citeauthoryear{{Beuermann}, {Kanbach}  \&
  {Berkhuijsen}}{{Beuermann} et~al.}{1985}]{Beuermann1985A&A...153...17B}
{Beuermann} K.,  {Kanbach} G.,   {Berkhuijsen} E.~M.,  1985, \aap, \href
  {http://adsabs.harvard.edu/abs/1985A%26A...153...17B} {153, 17}

\bibitem[\protect\citeauthoryear{{Caswell}}{{Caswell}}{1976}]{Caswell1976MNRAS.177..601C}
{Caswell} J.~L.,  1976, \mnras, \href
  {http://adsabs.harvard.edu/abs/1976MNRAS.177..601C} {177, 601}

\bibitem[\protect\citeauthoryear{{Caswell} \& {Haynes}}{{Caswell} \&
  {Haynes}}{1987}]{Caswell1987A&A...171..261C}
{Caswell} J.~L.,  {Haynes} R.~F.,  1987, \aap, \href
  {http://adsabs.harvard.edu/abs/1987A%26A...171..261C} {171, 261}

\bibitem[\protect\citeauthoryear{{Condon}}{{Condon}}{1992}]{Condon1992ARA&A..30..575C}
{Condon} J.~J.,  1992, \mn@doi [\araa] {10.1146/annurev.aa.30.090192.003043},
  \href {http://adsabs.harvard.edu/abs/1992ARA%26A..30..575C} {30, 575}

\bibitem[\protect\citeauthoryear{{Cordes} \& {Lazio}}{{Cordes} \&
  {Lazio}}{2002}]{Cordes2002astro.ph..7156C}
{Cordes} J.~M.,  {Lazio} T.~J.~W.,  2002, ArXiv Astrophysics e-prints, \href
  {http://adsabs.harvard.edu/abs/2002astro.ph..7156C} {}

\bibitem[\protect\citeauthoryear{{Deharveng} et~al.,}{{Deharveng}
  et~al.}{2010}]{Deharveng2010A&A...523A...6D}
{Deharveng} L.,  et~al., 2010, \mn@doi [\aap] {10.1051/0004-6361/201014422},
  \href {http://adsabs.harvard.edu/abs/2010A%26A...523A...6D} {523, A6}

\bibitem[\protect\citeauthoryear{{Epstein} \& {Feldman}}{{Epstein} \&
  {Feldman}}{1967}]{Epstein1967ApJ...150L.109E}
{Epstein} R.~I.,  {Feldman} P.~A.,  1967, \mn@doi [\apjl] {10.1086/180102},
  \href {http://adsabs.harvard.edu/abs/1967ApJ...150L.109E} {150, L109}

\bibitem[\protect\citeauthoryear{{Ferri{\`e}re}}{{Ferri{\`e}re}}{2001}]{Ferriere2001RvMP...73.1031F}
{Ferri{\`e}re} K.~M.,  2001, \mn@doi [Reviews of Modern Physics]
  {10.1103/RevModPhys.73.1031}, \href
  {http://adsabs.harvard.edu/abs/2001RvMP...73.1031F} {73, 1031}

\bibitem[\protect\citeauthoryear{{Fleishman} \& {Tokarev}}{{Fleishman} \&
  {Tokarev}}{1995}]{Fleishman1995A&A...293..565F}
{Fleishman} G.~D.,  {Tokarev} Y.~V.,  1995, \aap, \href
  {http://adsabs.harvard.edu/abs/1995A%26A...293..565F} {293}

\bibitem[\protect\citeauthoryear{{Garc{\'{\i}}a}, {Bronfman}, {Nyman}, {Dame}
  \& {Luna}}{{Garc{\'{\i}}a} et~al.}{2014}]{Garcia2014ApJS..212....2G}
{Garc{\'{\i}}a} P.,  {Bronfman} L.,  {Nyman} L.-{\AA}.,  {Dame} T.~M.,   {Luna}
  A.,  2014, \mn@doi [\apjs] {10.1088/0067-0049/212/1/2}, \href
  {http://adsabs.harvard.edu/abs/2014ApJS..212....2G} {212, 2}

\bibitem[\protect\citeauthoryear{{Green}}{{Green}}{2011}]{Green2011BASI...39..289G}
{Green} D.~A.,  2011, Bulletin of the Astronomical Society of India, \href
  {http://adsabs.harvard.edu/abs/2011BASI...39..289G} {39, 289}

\bibitem[\protect\citeauthoryear{{Green} et~al.,}{{Green}
  et~al.}{2015}]{Green2015ApJ...810...25G}
{Green} G.~M.,  et~al., 2015, \mn@doi [\apj] {10.1088/0004-637X/810/1/25},
  \href {http://adsabs.harvard.edu/abs/2015ApJ...810...25G} {810, 25}

\bibitem[\protect\citeauthoryear{{Guzm{\'a}n}, {May}, {Alvarez}  \&
  {Maeda}}{{Guzm{\'a}n} et~al.}{2011}]{Guzman2011A&A...525A.138G}
{Guzm{\'a}n} A.~E.,  {May} J.,  {Alvarez} H.,   {Maeda} K.,  2011, \mn@doi
  [\aap] {10.1051/0004-6361/200913628}, \href
  {http://adsabs.harvard.edu/abs/2011A%26A...525A.138G} {525, A138}

\bibitem[\protect\citeauthoryear{{Han}, {Manchester}, {Lyne}, {Qiao}  \& {van
  Straten}}{{Han} et~al.}{2006}]{Han2006ApJ...642..868H}
{Han} J.~L.,  {Manchester} R.~N.,  {Lyne} A.~G.,  {Qiao} G.~J.,   {van Straten}
  W.,  2006, \mn@doi [\apj] {10.1086/501444}, \href
  {http://adsabs.harvard.edu/abs/2006ApJ...642..868H} {642, 868}

\bibitem[\protect\citeauthoryear{{Haslam}, {Salter}, {Stoffel}  \&
  {Wilson}}{{Haslam} et~al.}{1982}]{Haslam1982A&AS...47....1H}
{Haslam} C.~G.~T.,  {Salter} C.~J.,  {Stoffel} H.,   {Wilson} W.~E.,  1982,
  \aaps, \href {http://adsabs.harvard.edu/abs/1982A%26AS...47....1H} {47, 1}

\bibitem[\protect\citeauthoryear{{Hindson} et~al.,}{{Hindson}
  et~al.}{2016}]{Hindson2016PASA...33...20H}
{Hindson} L.,  et~al., 2016, \mn@doi [\pasa] {10.1017/pasa.2016.19}, \href
  {http://adsabs.harvard.edu/abs/2016PASA...33...20H} {33, e020}

\bibitem[\protect\citeauthoryear{{Hou} \& {Han}}{{Hou} \&
  {Han}}{2014}]{Hou2014A&A...569A.125H}
{Hou} L.~G.,  {Han} J.~L.,  2014, \mn@doi [\aap] {10.1051/0004-6361/201424039},
  \href {http://adsabs.harvard.edu/abs/2014A%26A...569A.125H} {569, A125}

\bibitem[\protect\citeauthoryear{{Hurley-Walker} et~al.,}{{Hurley-Walker}
  et~al.}{2014}]{Hurley-Walker2014PASA...31...45H}
{Hurley-Walker} N.,  et~al., 2014, \mn@doi [\pasa] {10.1017/pasa.2014.40},
  \href {http://adsabs.harvard.edu/abs/2014PASA...31...45H} {31, 45}

\bibitem[\protect\citeauthoryear{{Jones} \& {Finlay}}{{Jones} \&
  {Finlay}}{1974}]{Jones1974AuJPh..27..687J}
{Jones} B.~B.,  {Finlay} E.~A.,  1974, Australian Journal of Physics, \href
  {http://adsabs.harvard.edu/abs/1974AuJPh..27..687J} {27, 687}

\bibitem[\protect\citeauthoryear{{Kassim}}{{Kassim}}{1987}]{Kassim1987PhDT........10K}
{Kassim} N.~E.~S.,  1987, PhD thesis, Maryland Univ., College Park.

\bibitem[\protect\citeauthoryear{{Krymkin}}{{Krymkin}}{1978}]{Krymkin1978Ap&SS..58..347K}
{Krymkin} V.~V.,  1978, \mn@doi [\apss] {10.1007/BF00644521}, \href
  {http://adsabs.harvard.edu/abs/1978Ap%26SS..58..347K} {58, 347}

\bibitem[\protect\citeauthoryear{{Kurtz}}{{Kurtz}}{2005}]{Kurtz2005IAUS..227..111K}
{Kurtz} S.,  2005, in {Cesaroni} R.,  {Felli} M.,  {Churchwell} E.,
  {Walmsley} M.,  eds,  IAU Symposium Vol. 227, Massive Star Birth: A
  Crossroads of Astrophysics. pp 111--119, \mn@doi{10.1017/S1743921305004424}

\bibitem[\protect\citeauthoryear{{Large}, {Mills}, {Little}, {Crawford}  \&
  {Sutton}}{{Large} et~al.}{1981}]{Large1981MNRAS.194..693L}
{Large} M.~I.,  {Mills} B.~Y.,  {Little} A.~G.,  {Crawford} D.~F.,   {Sutton}
  J.~M.,  1981, \mn@doi [\mnras] {10.1093/mnras/194.3.693}, \href
  {http://adsabs.harvard.edu/abs/1981MNRAS.194..693L} {194, 693}

\bibitem[\protect\citeauthoryear{{Large}, {Cram}  \& {Burgess}}{{Large}
  et~al.}{1991}]{Large1991Obs...111...72L}
{Large} M.~I.,  {Cram} L.~E.,   {Burgess} A.~M.,  1991, The Observatory, \href
  {http://adsabs.harvard.edu/abs/1991Obs...111...72L} {111, 72}

\bibitem[\protect\citeauthoryear{{Loi} et~al.,}{{Loi}
  et~al.}{2015}]{Loi2015MNRAS.453.2731L}
{Loi} S.~T.,  et~al., 2015, \mn@doi [\mnras] {10.1093/mnras/stv1808}, \href
  {http://adsabs.harvard.edu/abs/2015MNRAS.453.2731L} {453, 2731}

\bibitem[\protect\citeauthoryear{{McClure-Griffiths}, {Dickey}, {Gaensler}  \&
  {Green}}{{McClure-Griffiths}
  et~al.}{2004}]{McClure-Griffiths2004ApJ...607L.127M}
{McClure-Griffiths} N.~M.,  {Dickey} J.~M.,  {Gaensler} B.~M.,   {Green} A.~J.,
   2004, \mn@doi [\apjl] {10.1086/422031}, \href
  {http://adsabs.harvard.edu/abs/2004ApJ...607L.127M} {607, L127}

\bibitem[\protect\citeauthoryear{{Mezger} \& {Henderson}}{{Mezger} \&
  {Henderson}}{1967}]{Mezger1967ApJ...147..471M}
{Mezger} P.~G.,  {Henderson} A.~P.,  1967, \mn@doi [\apj] {10.1086/149030},
  \href {http://adsabs.harvard.edu/abs/1967ApJ...147..471M} {147, 471}

\bibitem[\protect\citeauthoryear{{Nord}, {Henning}, {Rand}, {Lazio}  \&
  {Kassim}}{{Nord} et~al.}{2006}]{Nord2006AJ....132..242N}
{Nord} M.~E.,  {Henning} P.~A.,  {Rand} R.~J.,  {Lazio} T.~J.~W.,   {Kassim}
  N.~E.,  2006, \mn@doi [\aj] {10.1086/504407}, \href
  {http://adsabs.harvard.edu/abs/2006AJ....132..242N} {132, 242}

\bibitem[\protect\citeauthoryear{{Orlando} \& {Strong}}{{Orlando} \&
  {Strong}}{2013}]{Orlando2013MNRAS.436.2127O}
{Orlando} E.,  {Strong} A.,  2013, \mn@doi [\mnras] {10.1093/mnras/stt1718},
  \href {http://adsabs.harvard.edu/abs/2013MNRAS.436.2127O} {436, 2127}

\bibitem[\protect\citeauthoryear{{Paladini}, {Davies}  \& {De
  Zotti}}{{Paladini} et~al.}{2004}]{Paladini2004MNRAS.347..237P}
{Paladini} R.,  {Davies} R.~D.,   {De Zotti} G.,  2004, \mn@doi [\mnras]
  {10.1111/j.1365-2966.2004.07210.x}, \href
  {http://adsabs.harvard.edu/abs/2004MNRAS.347..237P} {347, 237}

\bibitem[\protect\citeauthoryear{{Peterson} \& {Webber}}{{Peterson} \&
  {Webber}}{2002}]{Peterson2002ApJ...575..217P}
{Peterson} J.~D.,  {Webber} W.~R.,  2002, \mn@doi [\apj] {10.1086/341258},
  \href {http://adsabs.harvard.edu/abs/2002ApJ...575..217P} {575, 217}

\bibitem[\protect\citeauthoryear{{Quireza}, {Rood}, {Bania}, {Balser}  \&
  {Maciel}}{{Quireza} et~al.}{2006}]{Quireza2006ApJ...653.1226Q}
{Quireza} C.,  {Rood} R.~T.,  {Bania} T.~M.,  {Balser} D.~S.,   {Maciel} W.~J.,
   2006, \mn@doi [\apj] {10.1086/508803}, \href
  {http://adsabs.harvard.edu/abs/2006ApJ...653.1226Q} {653, 1226}

\bibitem[\protect\citeauthoryear{{Reid} et~al.,}{{Reid}
  et~al.}{2014}]{Reid2014ApJ...783..130R}
{Reid} M.~J.,  et~al., 2014, \mn@doi [\apj] {10.1088/0004-637X/783/2/130},
  \href {http://adsabs.harvard.edu/abs/2014ApJ...783..130R} {783, 130}

\bibitem[\protect\citeauthoryear{{Roger}, {Costain}, {Landecker}  \&
  {Swerdlyk}}{{Roger} et~al.}{1999}]{Roger1999A&AS..137....7R}
{Roger} R.~S.,  {Costain} C.~H.,  {Landecker} T.~L.,   {Swerdlyk} C.~M.,  1999,
  \mn@doi [\aaps] {10.1051/aas:1999239}, \href
  {http://adsabs.harvard.edu/abs/1999A%26AS..137....7R} {137, 7}

\bibitem[\protect\citeauthoryear{{Scheuer} \& {Ryle}}{{Scheuer} \&
  {Ryle}}{1953}]{Scheuer1953MNRAS.113....3S}
{Scheuer} P.~A.~G.,  {Ryle} M.,  1953, \mnras, \href
  {http://adsabs.harvard.edu/abs/1953MNRAS.113....3S} {113, 3}

\bibitem[\protect\citeauthoryear{{Sun} \& {Reich}}{{Sun} \&
  {Reich}}{2010}]{Sun2010RAA....10.1287S}
{Sun} X.-H.,  {Reich} W.,  2010, \mn@doi [Research in Astronomy and
  Astrophysics] {10.1088/1674-4527/10/12/009}, \href
  {http://adsabs.harvard.edu/abs/2010RAA....10.1287S} {10, 1287}

\bibitem[\protect\citeauthoryear{{Sun} \& {Reich}}{{Sun} \&
  {Reich}}{2012}]{Sun2012A&A...543A.127S}
{Sun} X.~H.,  {Reich} W.,  2012, \mn@doi [\aap] {10.1051/0004-6361/201218802},
  \href {http://adsabs.harvard.edu/abs/2012A%26A...543A.127S} {543, A127}

\bibitem[\protect\citeauthoryear{{Sun}, {Reich}, {Waelkens}  \&
  {En{\ss}lin}}{{Sun} et~al.}{2008}]{Sun2008A&A...477..573S}
{Sun} X.~H.,  {Reich} W.,  {Waelkens} A.,   {En{\ss}lin} T.~A.,  2008, \mn@doi
  [\aap] {10.1051/0004-6361:20078671}, \href
  {http://adsabs.harvard.edu/abs/2008A%26A...477..573S} {477, 573}

\bibitem[\protect\citeauthoryear{{Tingay} et~al.,}{{Tingay}
  et~al.}{2013}]{Tingay2013PASA...30....7T}
{Tingay} S.~J.,  et~al., 2013, \mn@doi [\pasa] {10.1017/pasa.2012.007}, \href
  {http://adsabs.harvard.edu/abs/2013PASA...30....7T} {30, 7}

\bibitem[\protect\citeauthoryear{{Wayth} et~al.,}{{Wayth}
  et~al.}{2015}]{Wayth2015PASA...32...25W}
{Wayth} R.~B.,  et~al., 2015, \mn@doi [\pasa] {10.1017/pasa.2015.26}, \href
  {http://adsabs.harvard.edu/abs/2015PASA...32...25W} {32, e025}

\bibitem[\protect\citeauthoryear{{Westfold}}{{Westfold}}{1959}]{Westfold1959ApJ...130..241W}
{Westfold} K.~C.,  1959, \mn@doi [\apj] {10.1086/146713}, \href
  {http://adsabs.harvard.edu/abs/1959ApJ...130..241W} {130, 241}

\bibitem[\protect\citeauthoryear{{Wright} et~al.,}{{Wright}
  et~al.}{2010}]{Wright2010AJ....140.1868W}
{Wright} E.~L.,  et~al., 2010, \mn@doi [\aj] {10.1088/0004-6256/140/6/1868},
  \href {http://adsabs.harvard.edu/abs/2010AJ....140.1868W} {140, 1868}

\bibitem[\protect\citeauthoryear{{Zheng} et~al.,}{{Zheng}
  et~al.}{2016}]{Zheng2016MNRAS.tmp.1510Z}
{Zheng} H.,  et~al., 2016, \mn@doi [\mnras] {10.1093/mnras/stw2525}, \href
  {http://adsabs.harvard.edu/abs/2016MNRAS.tmp.1510Z} {}

\bibitem[\protect\citeauthoryear{{de Oliveira-Costa}, {Tegmark}, {Gaensler},
  {Jonas}, {Landecker}  \& {Reich}}{{de Oliveira-Costa}
  et~al.}{2008}]{Oliveira-Costa2008MNRAS.388..247D}
{de Oliveira-Costa} A.,  {Tegmark} M.,  {Gaensler} B.~M.,  {Jonas} J.,
  {Landecker} T.~L.,   {Reich} P.,  2008, \mn@doi [\mnras]
  {10.1111/j.1365-2966.2008.13376.x}, \href
  {http://adsabs.harvard.edu/abs/2008MNRAS.388..247D} {388, 247}

\makeatother
\end{thebibliography}

%%%%%%%%%%%%%%%%%%%%%%%%%%%%%%%%%%%%%%%%%%%%%%%%%%

%%%%%%%%%%%%%%%%% APPENDICES %%%%%%%%%%%%%%%%%%%%%

%\appendix

%%%%%%%%%%%%%%%%%%%%%%%%%%%%%%%%%%%%%%%%%%%%%%%%%%

% Don't change these lines
\bsp	% typesetting comment
\label{lastpage}
\end{document}